\documentclass{aa}

\usepackage{multirow}
\usepackage{booktabs}
\usepackage{graphicx}
\usepackage{menukeys}
\usepackage{algorithm}
\usepackage[noend]{algpseudocode}
\usepackage[colorlinks=true, allcolors=blue]{hyperref}
\usepackage{txfonts}
\newcommand{\molh}{H$_{2}$ } 
\newcommand{\cii}{[C{\small II}] } 

\usepackage{cuted}
\makeatletter
\renewcommand*\aa@pageof{, page \thepage{} of \pageref*{LastPage}}
\makeatother
\begin{document} 

   \title{Fingerprinting the effects of hyperfine structure on CH and OH far infrared spectra using Wiener filter deconvolution\thanks{The reduced spectra (Fig.~\ref{fig:sofia_spec}) are only available at the CDS via anonymous ftp to \url{cdsarc.u-strasbg.fr} (\url{130.79.128.5}) or via \url{http://cdsweb.u-strasbg.fr/cgi-bin/qcat?J/A+A/}}}

   \author{Arshia M. Jacob
          \inst{1}
          \and
          Karl M. Menten\inst{1}
          \and
          Helmut Wiesemeyer\inst{1}
          \and
          Min-Young Lee\inst{1,2}
          \and 
          Rolf G\"{u}sten\inst{1}
          \and 
          Carlos A. Dur\'{a}n\inst{1}
          }

   \institute{Max-Planck-Institut f\"{u}r Radioastronomie, Auf dem H\"{u}gel 69, 53121 Bonn, Germany
   \and 
   Korea Astronomy and Space Science Institute, 776 Daedeokdae-ro, 34055 Daejeon, Republic of Korea
   \\
   \email{ajacob@mpifr-bonn.mpg.de}}

   \date{Received September 15, 1996; accepted March 16, 1997}
  \titlerunning{Fingerprinting the effects of hyperfine structure}
   \authorrunning{A. Jacob et al.}
% \abstract{}{}{}{}{} 
% 5 {} token are mandatory
 
  \abstract
   { Despite being a commonly observed feature, the modification of the velocity structure in spectral line profiles by hyperfine structure complicates the interpretation of spectroscopic data. This is particularly true for observations of simple molecules such as CH and OH toward the inner Galaxy, which show a great deal of velocity crowding.
      } 
   {In this paper, we investigate the influence of hyperfine splitting on complex spectral lines, with the aim of evaluating canonical abundances by decomposing their dependence on hyperfine structures. This is achieved from first principles through deconvolution. }
   { We present high spectral resolution observations of the rotational ground state transitions of CH near 2$\,$THz seen in absorption toward the strong FIR-continuum sources AGAL010.62$-00$.384, AGAL034.258+00.154, AGAL327.293$-00$.579, AGAL330.954$-00$.182, AGAL332.826$-00$.549, AGAL351.581$-00$.352 and SgrB2(M). These were observed with the GREAT instrument on board SOFIA. The observed line profiles of CH were deconvolved from the imprint left by the lines' hyperfine structures using the Wiener filter deconvolution, an optimised kernel acting on direct deconvolution. 
   }
   {The quantitative analysis of the deconvolved spectra first entails the computation of CH column densities. Reliable $N$(CH) values are of importance owing to the status of CH as a powerful tracer for \molh in the diffuse regions of the interstellar medium. The $N$(OH)/$N$(CH) column density ratio is found to vary within an order of magnitude with values ranging from one to 10, for the individual sources that are located outside the Galactic centre. Using CH as a surrogate for H$_{2}$, we determined the abundance of the OH molecule to be $X$(OH) = $1.09\times10^{-7}$ with respect to H$_{2}$. The radial distribution of CH column densities along the sightlines probed in this study, excluding SgrB2(M), showcase a dual peaked distribution peaking between 5 and 7~kpc. The similarity between the correspondingly derived column density profile of H$_{2}$ with that of the CO-dark \molh gas traced by the cold neutral medium component of \cii 158$\, \mu$m emission across the Galactic plane, further emphasises the use of CH as a tracer for H$_{2}$.}
   {}

 
   \maketitle
   
%
%________________________________________________________________

\section{Introduction}\label{sec:Introduction}

\begin{figure}[ht]
\includegraphics[width=9cm]{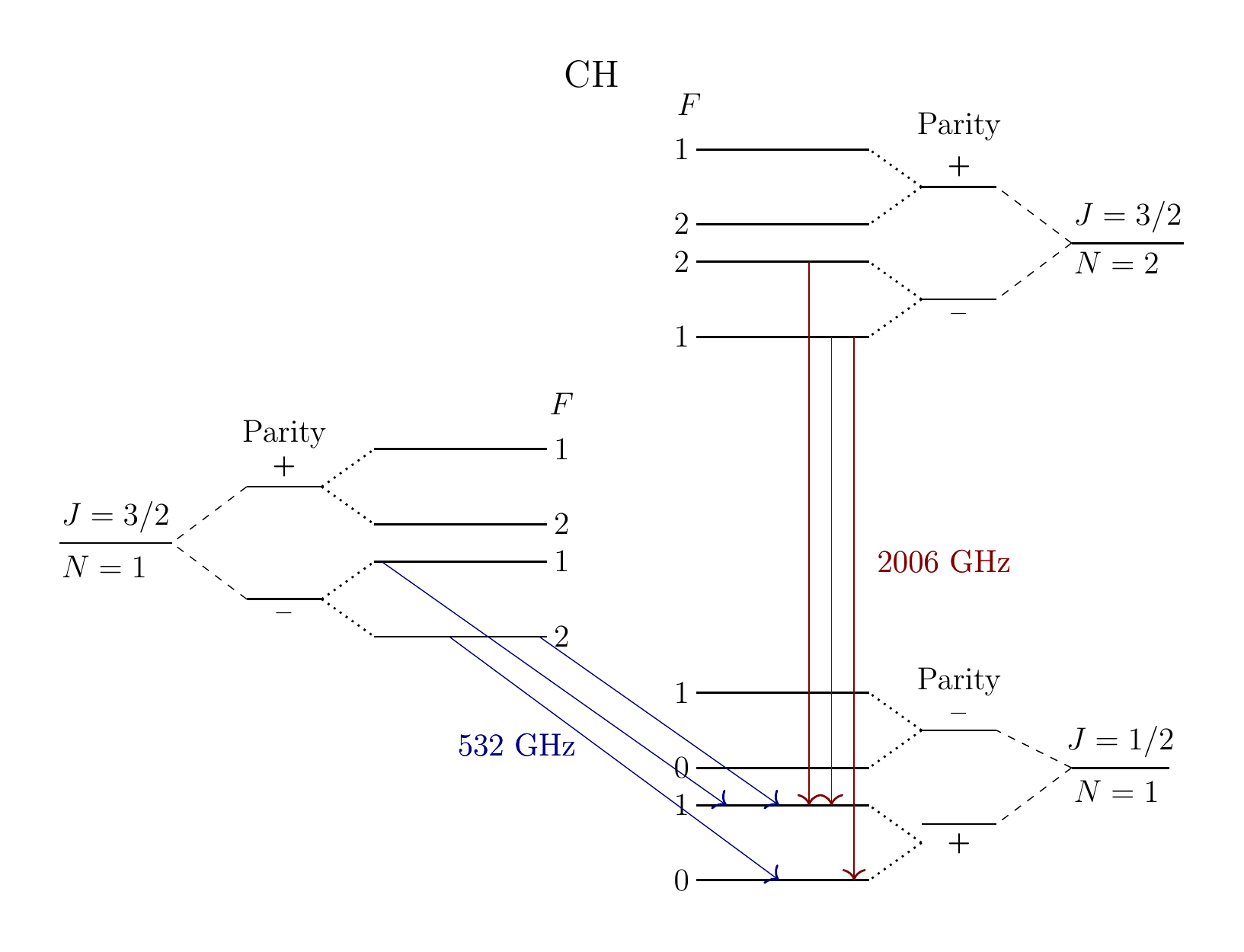}
\caption{Low-lying energy level structure of CH. The $N, J = 2,3/2 \rightarrow 1,1/2$ HFS transitions of CH near $2006\,$GHz that were observed using SOFIA are represented in red while the $N, J = 1,3/2 \rightarrow 1,1/2$ HFS transitions near 532~GHz observed with \textit{Herschel}/HIFI are displayed in blue. }
\label{fig:energy_level}
\end{figure}

Hyperfine structure (HFS) splitting is a commonly observed phenomenon in spectroscopy that arises from interactions between unfilled electron shells and the nuclear spin magnetic moment of atoms or molecules. It leaves a unique imprint on the spectrum of a given chemical species and transition, much like fingerprints. Over the years, knowledge of HFS in atoms and molecules has served both atomic-, molecular-, and astro-physicists alike in interpreting observations. The effects of such splitting, despite being small, leads to energy differences that are typically of the order of $\sim0.03\,$K in terms of temperature. Since HFS interactions broaden, shift, and even alter the shape of spectra \citep{booth1983effect}, not taking them into consideration might lead to erroneous interpretations of astrophysical quantities. Overall, taking HFS splitting into account makes it possible to precisely model and resolve the underlying spectral line shapes, since the observed integral line profile is a superposition of the various HFS components. In recent years, the onset of improved observational techniques with greater sensitivity and higher spectral resolution have paved the way for rigorous spectroscopic studies and abundance measurements, in particular in the sub-millimetre and far-infrared (FIR) regions. Observations made between 2010 and 2013 with the Herschel-High Frequency Instrument for the Far-Infrared \citep[HIFI,][]{de2010herschel}, later followed by the German REceiver at Terahertz frequencies \citep[GREAT,][]{heyminck2012great} on board of the Stratospheric Observatory for Infrared Astronomy \citep[SOFIA,][]{young2012early} have resulted in high resolution (down to $10^{-7}$) absorption spectra of a plethora of light hydride molecules of which the study is important for various reasons \citep[for an overview see][]{Gerin2016}. In many cases, the background sources reside far away in the inner Galaxy. Thus the differential Galactic rotation causes complex velocity crowding, which is further complicated by the HFS of many of these molecules, including CH and OH, which are studied here. All this requires advanced techniques of analysis of the observed line profiles. Several different approaches have previously been used to fit the HFS of various spectral lines through techniques ranging from simulated annealing algorithms \citep{wiesemeyer2016far} and Monte Carlo approaches that use genetic algorithms \citep{estalella2017HFS} to direct deconvolution \citep{gerin2010interstellar}, and multi-Gaussian fitting techniques \citep{neufeld2015herschel,persson2016ortho}. In this paper, we introduce a technique that decomposes the contributions of HFS from astrophysical spectra, based on the theory developed by \citet{wiener1949extrapolation}. A linear and time-invariant method, the Wiener filter theory has found its application in practical signal reconstruction scenarios. It stands out from other filtering techniques because it provides the maximum a posteriori estimate of the unknown signal through a simple least squares analysis. Consequently, the Wiener filter has also found its application in astronomy as a tool that aids in cleaning contaminated images retrieved by aperture synthesis techniques \citep{Caprari:00}. In cosmological studies, deconvolution using the Wiener filter has been employed to accurately decompose the observed polarization of the cosmic microwave background (CMB) radiation into maps of pure polarization for its two components: E and B  \citep{bunn2017pure}. Given the proven capability of the Wiener filter deconvolution as a method of choice for decomposition, we decided to adapt it for a novel application, the deconvolution of HFS from astrophysical spectra.

Concentrating for most of this paper on the transitions of a single molecule, we illustrate the working of the Wiener filter algorithm on spectra that are of astrophysical importance, through the spectroscopy of the $N, J~=~2, 3/2~\rightarrow ~1, 1/2$ ground state rotational transitions of CH, the methylidene radical. CH was amongst the first molecules detected in the interstellar medium (ISM) by \citet{dunham1937interstellar}, and identified by \citet{swings1937considerations}. Being the simplest organic carbyne and one of the lightest hydrides, CH initiates the formation of a large fraction of the molecules present in the ISM, thereby playing a pivotal role in both the physics and chemistry of interstellar gas clouds. CH has been widely studied in various spectral regimes to infer the thermal, chemical and dynamical evolution of diffuse and translucent cloud regions \citep{rydbeck1973radio,turner1974microwave,stacey1987detection}. Moreover, the advent of space- and air-borne telescopes like \textit{Herschel} and SOFIA have established the use of CH alongside other interstellar hydrides as tracers for \molh in the diffuse regions of the ISM \citep{gerin2010interstellar, wiesemeyer2018unveiling}. Figure~\ref{fig:energy_level} shows the low-lying energy level structure of the ground electronic state of the CH molecule, abiding by Hund's case b. The total angular momentum states labelled $J$ splits into a pair of $\Lambda$-doublet levels of opposing parity ($\pm\Lambda$). Each of these are further split into hyperfine levels given by $F = J \pm I$, where $I$ is the nuclear spin of the H atom, I($H$) = 1/2.

The outline of this paper is as follows: In Sect.~\ref{subsec:WFD}, we develop the theory behind the working of the Wiener filter, and in Sect.~\ref{sec:WF_algo}, we detail our procedure for spectral line analysis: the Wiener filter algorithm. Having motivated our astronomical case study on the $N,J = 2,3/2 \rightarrow 1,1/2$ HFS transitions of CH, in Sect.~\ref{subsec:astro_app}, we describe our observations. In Sect.~\ref{sec:analysis}, we present the implementation of the Wiener filter on the absorption spectra and its analysis, while the results obtained are discussed in Sect.~\ref{sec:results}, in which we also compare the distribution of CH with that of OH, and explore the usefulness of CH as a tracer for H$_{2}$. We conclude this work in Sect.\ref{sec:summary}. 

%

%__________________________________________________________________

\section{Theory}\label{sec:theory}
Convolution is a formal mathematical operation used to describe the interaction between an input signal, $f(\nu)$, and an impulse response function, $h(\nu)$, to produce an output signal, $g(\nu)$, in a linear system. In spectroscopic terms, the input and output signals are analogous to the original and observed spectra while the impulse response in this case arises from the HFS splitting. For a discrete sample, such a convolution model can be mathematically formulated as follows:
\begin{equation}
    g(\nu) = f(\nu)\circledast h(\nu) = \sum_{\nu^{\prime} = -\infty}^{+\infty} f(\nu^{\prime}) h(\nu - \nu^{\prime}) \, .
    \label{eqn:convolution}
\end{equation}
 With astrophysical observations, one is posed with the inverse problem of convolution, which is deconvolution, wherein the output spectrum is known, while its main constituent, the original spectrum, remains unknown. 
The convolution theorem renders this task trivial, as it reduces the operations of both convolution and deconvolution into those of simple multiplication and division, in Fourier space, given a priori knowledge of $g(\nu)$ and $h(\nu)$. Hence, in Fourier space, Eq.~\ref{eqn:convolution} becomes
\begin{align}
G(\tilde{\nu}) &= F(\tilde{\nu}).H(\tilde{\nu}) \nonumber \quad \text{Convolution;}\\
F(\tilde{\nu}) &= \frac{G(\tilde{\nu})}{H(\tilde{\nu})} \quad \text{Deconvolution} \, ,
\end{align}
where $G(\tilde{\nu})$, $F(\tilde{\nu})$ and $H(\tilde{\nu})$ represent the Fourier transforms (FTs) of $g(\nu)$, $f(\nu)$ and $h(\nu)$, respectively. However, such a point-wise division of the Fourier transforms is sufficient to obtain an estimate of $f(\nu)$, only as long as the transform of the response function remains non-zero over all frequencies. Moreover, realistic systems are influenced by noise, $n(\nu)$, which further degrades the observed spectrum, $d(\nu)$. 
This additive noise is often amplified when directly deconvolved, because it acts as a low-pass filter and gives rise to faux features in the reconstructed signal. Such issues of sensitivity are often tackled by using filters, and in the following sections, we introduce and employ one such filter, namely the Wiener filter.  
\subsection{Wiener filter deconvolution}\label{subsec:WFD}
Wiener's theory formulated a linear tool, the Wiener filter (hereafter WF) for additive noise reduction aimed to resolve the signal restoration problems (singularities) faced by direct deconvolution. The WF model assumes all signals to be stationary\footnote{Time invariant in first and second order statistics.}, and modelled by linear stochastic processes with a signal-independent noise. The WF output in the Fourier or inverse frequency domain, $\hat{F}(\tilde{\nu})$ is the product of the (noise) degraded observed spectrum, $D(\tilde{\nu})$, and the frequency response of the filter, $W(\tilde{\nu})$: \\
\begin{equation}
\hat{F}(\tilde{\nu}) = D(\tilde{\nu}) W(\tilde{\nu}) \quad \Rightarrow \quad \hat{F}(\tilde{\nu})= \left[ F(\tilde{\nu}) H(\tilde{\nu}) + N(\tilde{\nu}) \right] W(\tilde{\nu})
\end{equation}

\begin{figure}
    \centering
    \includegraphics[width=0.5\textwidth]{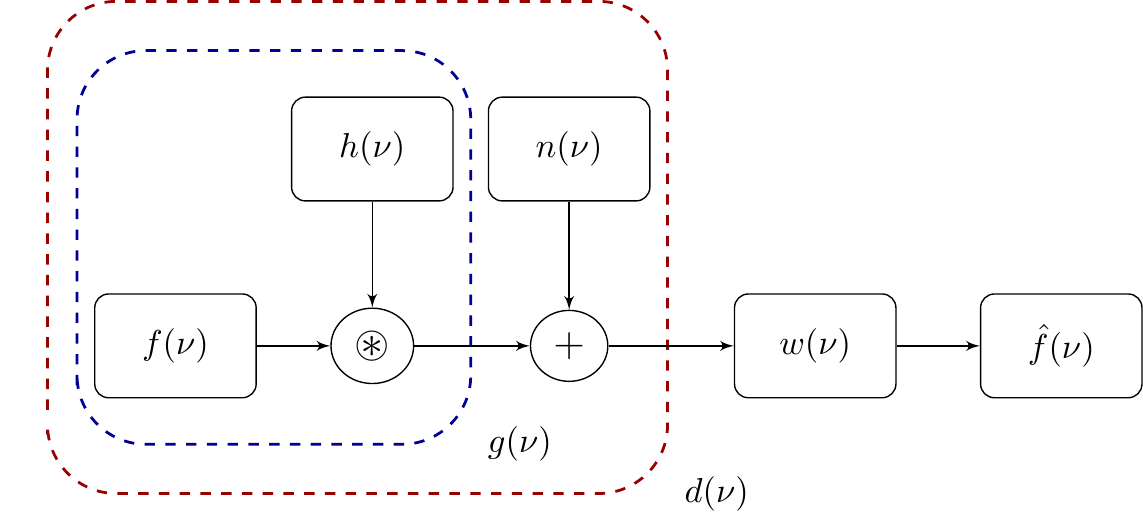}
    \caption{Schematic diagram showcasing working principle of the Wiener filter in frequency space.}
    \label{fig:flow_chart}
\end{figure}

The filter further models the input signal by implementing a minimum mean square error (MSE) $\epsilon^2$ constraint on the deconvolution:
\begin{align}
\epsilon^2 &= \sum_{\tilde{\nu}} \left| F(\tilde{\nu}) - \hat{F}(\tilde{\nu}) \right|^2\\
           &= \sum_{\tilde{\nu}} \left| F(\tilde{\nu}) - \left[F(\tilde{\nu}) H(\tilde{\nu}) + N(\tilde{\nu})\right] W(\tilde{\nu}) \right|^2\\
           &=  \sum_{\tilde{\nu}} \left| F(\tilde{\nu})[1- H(\tilde{\nu})W(\tilde{\nu})] - N(\tilde{\nu})W(\tilde{\nu})\right|^2 \label{eqn:quadraticterm}
\end{align}
Expanding the quadratic term in Eq.~\ref{eqn:quadraticterm},
\begin{align}
\begin{split}
\epsilon^2 &= \sum_{\tilde{\nu}} \left( \left[1-H(\tilde{\nu})W(\tilde{\nu})\right]\left[1-H(\tilde{\nu})W(\tilde{\nu})\right]^{*}\right)|F(\tilde{\nu})|^2 \\
&\quad- \left[1-H(\tilde{\nu})W(\tilde{\nu})\right]W(\tilde{\nu})^{*}\left(F(\tilde{\nu})N(\tilde{\nu})^{*} \right)\\
&\quad-  W(\tilde{\nu})\left[1-H(\tilde{\nu})W(\tilde{\nu})\right]^{*}\left(N(\tilde{\nu})F(\tilde{\nu})^{*} \right) + |W(\tilde{\nu})|^2|N(\tilde{\nu})|^2 \, . 
\label{eqn:expansion}
\end{split}
\end{align}
We assume that $N(\tilde{\nu})$ is independent of $F(\tilde{\nu})$, therefore $ \left(F(\tilde{\nu}).N(\tilde{\nu})^{*}\right) =  \left(N(\tilde{\nu}).F(\tilde{\nu})^{*} \right) = 0$. This reduces Eq.~\ref{eqn:expansion} to,
\begin{align}
\epsilon^2 &= \sum_{\tilde{\nu}} \left| F(\tilde{\nu}) \right|^2 \left| 1 - H(\tilde{\nu})W(\tilde{\nu})\right|^2 + \left| {N(\tilde{\nu})} \right|^2 \left| W(\tilde{\nu}) \right|^2 \, .
\end{align}
The derivative of the MSE, $\epsilon^2$ with respect to $W(\tilde{\nu})$ yields
\begin{align}
\frac{\partial \epsilon^2}{\partial W(\tilde{\nu})} &= \left | F(\tilde{\nu}) \right|^2 \left[ 2(1 - W^*H^*)(-H)\right] + \left| N(\tilde{\nu}) \right|^2 \left[ 2W^*\right] \, ,
\label{eqn:min_errors}
\end{align}
using $\frac{\partial(zz^*)}{\partial z} = 2z^*$ (property of conjugates), where the asterisk represents the complex conjugate. Minimising $\partial \epsilon^2/\partial W(\tilde{\nu})$ by equating Eq.~\ref{eqn:min_errors} to zero and solving for $W(\tilde{\nu})$ results in the general form of the Wiener filter,
\begin{equation*}
W(\tilde{\nu}) = \frac{H^*(\tilde{\nu})}{ \left| H(\tilde{\nu}) \right|^2 + \left| \frac{N(\tilde{\nu})}{F(\tilde{\nu})} \right|^2 } \, .
\end{equation*}
Since the exact forms of both $N(\tilde{\nu})$ and $F(\tilde{\nu})$ are model-dependent, they remain as unknowns for most practical systems.  However, for additive white noise that is independent of the signal, the $\frac{F(\tilde{\nu})}{N(\tilde{\nu})}$ ratio can be approximated by the signal-to-noise ratio (S/N), or some form of normalized noise variance, $\sigma_{\tilde{\nu}}$, characterised by the system noise. 
Rearranging the terms to fit the inverse filter formulation, the Wiener filter and restored spectra are given as: 
\begin{align}
W(\tilde{\nu}) &= \frac{1}{H(\tilde{\nu})} \frac{\left| H(\tilde{\nu}) \right|^2}{ \left| H(\tilde{\nu}) \right|^2 + \left| \frac{1}{\text{S/N}} \right|^2 } \, \, \quad \, \, \text{and}
\label{eqn:WF}
\end{align}

\begin{align}
\hat{F}(\tilde{\nu}) &= \underbrace{\frac{D(\tilde{\nu})}{H(\tilde{\nu})}}_\text{Inverse filter} \underbrace{\frac{\left| H(\tilde{\nu}) \right|^2}{ \left| H(\tilde{\nu}) \right|^2 + \left| \frac{1}{\text{S/N}} \right|^2 }}_\text{Kernel} \, .
\label{eqn:signal_recovery}
\end{align}
Figure~\ref{fig:flow_chart} illustrates the idea behind the WF deconvolution algorithm as expressed in Eq.~\ref{eqn:signal_recovery}. Therefore, a reliable estimation of the "original", underlying spectrum using the WF deconvolution is dependent upon the assumptions made in determining the kernel term, $W(\tilde{\nu})$, of the filter (Eq.~\ref{eqn:signal_recovery}). The kernel term is dependent on the noise of the system and on the response function, which in this case is tailored to address the effects of HFS splitting. The following section details the kernel estimation, as well as the adopted algorithm.  

\subsection{WF algorithm}\label{sec:WF_algo}
This section features a brief description of our WF fitting and deconvolution algorithm written in python, based on the mathematical formulation discussed in ~Sect.~\ref{subsec:WFD}. The WF algorithm uses the radiative transfer equation to express the observed line profile, $T_\text{l}$, of a given molecular transition containing $n_\text{HFS}$ hyperfine components in terms of optical depths, $\tau$. This is done under the assumption of local thermodynamic equilibrium (LTE) and unsaturated absorption. The formal solution of the radiative transfer equation for a constant excitation temperature reads:
\begin{equation}
    T_{\text{l}} = \Phi_\text{l}T_\text{ex}(1-\text{e}^{-\tau}) +  \Phi_\text{c}T_{\text{c}}\text{e}^{-\tau},
    \label{eqn:radtrans}
\end{equation}
where the excitation temperature, $T_\text{ex}$ and the background continuum, $T_\text{c}$ are expressed as Rayleigh-Jeans equivalent temperatures, and $\Phi_\text{l}$ and $\Phi_\text{c}$ are the beam filling factors of the line and continuum emission, respectively. For the specific case of absorption spectroscopy of far-infrared ground state transitions in low-excitation gas ($h\nu$ >> $kT_{ex}$), we can neglect the first term and recast the radiative transfer equation displayed in Eq.~\ref{eqn:radtrans} as $T_{\text{l}} = \Phi_\text{c}T_{\text{c}}\text{e}^{-\tau}$. In cases where we cannot neglect the emission term, particularly in the immediate environment of the continuum source, a realistic description needs to account for an excitation gradient, and therefore requires a full non-LTE solution, which is outside the scope of this paper. 

The next step of the algorithm determines the kernel term, which constitutes the hyperfine response and the noise term. Hence, the quality of the restored spectrum relies both on the impulse response of the HFS and an accurate representation of the system noise. The effects of HFS splitting are characterised by using weighted impulses in the frequency domain. The impulse response of each hyperfine component is represented by a Dirac-$\delta$ function, weighted by the relevant spectroscopic parameters.
The weighting function, $\Omega_\text{(HFS)}$, accounts for the fraction of the specified species in the upper energy level, $E_{u}$ (in Kelvin) of a given transition that has an upper-level degeneracy, $g_{u}$ and Einstein A coefficient, $A_{\text{E}}$ as follows:
\begin{equation}
\Omega_\text{(HFS)} = \frac{g_{u} A_{\text{E}}}{Q(T_\text{rot})}\text{exp}\left( \frac{-E_{u}}{T_\text{ex}}\right) \, .
\label{eqn:HFS_weights}
\end{equation}
For a given hyperfine transition, all of the above spectroscopic terms remain constant, except for the partition function, $Q$, which itself is a function of rotation temperature, $T_\text{rot}$, which in our case (LTE) is equal to the excitation temperature, $T_\text{ex}$. When deriving the formulation of the WF, in Sect.~\ref{subsec:WFD}, the noise term was approximated to be the inverse of the S/N. While this serves as an acceptable choice for the noise contribution, it can be further refined and expressed explicitly as a function of the receiver system, which we will not discuss here in this paper. 

The WF kernel is then determined following Eq.~\ref{eqn:WF}, and the algorithm then fits the spectrum. The standard deviation of the residual distribution is used to measure the quality of the spectral fit, and based on this, the algorithm carries out deconvolution following Eq.~\ref{eqn:signal_recovery}. In the last step, the WF algorithm applies an inverse FT to convert the modelled spectrum $\hat{F}(\tilde{\nu})$ into $\hat{f}(\nu)$ in frequency space. The deconvolved optical depth signature, $\tau_{decon}$ is then converted to molecular column density values using
\begin{equation}
\left( \frac{\text{d}N}{\text{d}\varv} \right)_{i} = \frac{8 \pi \nu_{i}^3}{c^3\Omega}\left[\text{exp}\left(\frac{h \nu_{i}}{k_{\text{B}}T_{\text{ex}}}\right) - 1 \right]^{-1} (\tau_{i})_{decon} \, ,
\label{eq:coldens}
\end{equation}
for each velocity channel, $i$, since, the backend provides the frequencies (and the velocities) as a discrete sequence of data, which was further smoothed by us to a useful spectral resolution as explained in Sect.~\ref{subsec:astro_app} and where $\Omega$ represents the mean HFS weight. Having established the general scheme of the WF algorithm, the following sections detail its functioning on spectra that are of astrophysical importance.

\section{GREAT observations and data reduction}\label{subsec:astro_app}
The observations presented here were performed using upGREAT\footnote{upGREAT, an extension of the German REceiver for Astronomy at Terahertz frequencies, is a development by the MPI f\"{u}r Radioastronomie and KOSMA/Universit\"{a}t zu K\"{o}ln, in cooperation with the MPI f\"{u}r Sonnensystemforschung, and the DLR Institut f\"{u}r Optische Sensorsysteme.}/LFA ~\citep{2018JAI.....740014R}, on board SOFIA over its cycle 5 and 6 campaigns, in a number of the observatory's flights (on 26 June and 6 July in 2017, and 20 and 22 June in 2018). Our source selection consists of several far-infrared bright hot cores from the ATLASGAL\footnote{APEX Telescope Large Area Survey of the GALaxy (ATLASGAL)} survey \citep{csengeri2017atlasgal} and SgrB2(M). The observed sources, whose strong far-infrared dust continuum emission serves as background for our absorption measurements, are listed in Table.~\ref{tab:sources}. \\
\begin{table*}
\caption{Continuum source parameters. }
\begin{center}

\begin{tabular}{cccccccc}
\hline \hline
Continuum & R.A.~(J2000) & Dec.~(J2000) & $l$ & $b$ & D & $\varv_\text{LSR}$ & $T_\text{c}$ \\
Source & [$^{\text{h} \, \text{m} \, \text{s}}$] & [$^{\circ} \, ^{\prime} \, ^{\prime \prime}$] & [${}^{\circ}$] & [${}^{\circ}$] & [kpc] & [km~s$^{-1}$] & [K] \\
\hline
%W49$\,$N & 43.166 & 0.012 & 11.4  & +11 & 17.0\\
%W51$\,$e1/e2 & 49.489 & -0.388 & 5.4 & +56 & 9.2\\
AGAL010.624$-$00.384  (G10.62)& 18:10:28.69 & $-$19:55:50.0 &  $\,\,\,$10.624  & $-$0.383 & 4.9 & $\,\,-$2.9  & $\,\,$8.2 \\
AGAL034.258+00.154  (G34.26) & 18:53:18.49 & +01:14:58.7 & $\,\,\,$34.257  & +0.153 & 1.6 & +58.5  & $\,\,$7.5 \\
AGAL327.293$-$00.579 (G327.29)& 15:53:08.55 & $-$54:37:05.1 & 327.292 & $-$0.578 & 3.1 & $-$44.7 & $\,\,$2.5 \\
AGAL330.954$-$00.182 (G330.95) &  16:09:53.01 & $-$51:54:55.0 & 330.954 & $-$0.183 & 9.3 & $-$91.2 & 11.7 \\
AGAL332.826$-$00.549 (G332.83) & 16:20:10.65 & $-$50:53:17.6 & 332.825 & $-$0.549 & 3.6 & $-$57.1 & $\,\,$7.7 \\
AGAL351.581$-$00.352 (G351.58) & 17:25:25.03 & $-$36:12:45.3 & 351.580 & $-$0.352 & 6.8 & $-$95.9 & $\,\,$5.2 \\
Sgr$\,$B2(M) & 17:47:20.16 & $-$28:23:04.5&  $\quad$0.667  & $-$0.036 & 8.3 & +64.0 & 15.1 \\

\hline 
\label{tab:sources}
\end{tabular}
\tablefoot{Columns are, left to right, source designation, equatorial coordinates, Galactic coordinates, heliocentric distance, LSR velocity, and signal band continuum brightness temperature derived by means of a DSB calibration. Heliocentric distance references: AGAL010.624$-$00.384:~\citet{sanna2014trigonometric}, AGAL034.258+00.154:~\citet{zhang2009trigonometric}, AGAL327.293$-$00.579:~\citet{urquhart2013rms}, AGAL330.954$-$00.182:~\citet{urquhart2012rms},  AGAL332.826$-$00.549:~\citet{moises2011spectrophotometric}, AGAL351.581$-$00.352:~\citet{10.1111/j.1365-2966.2011.19418.x}, Sgr~B2(M):~\citet{reid2014trigonometric}.}
\end{center}
\end{table*}
The receiver configuration is comprised of the (7+7) pixel low frequency array (LFA) receiver on upGREAT, in dual polarization. Only one set of HFS transitions near $2006\,$GHz (Table.~\ref{tab:freq}) was observed, because the atmospheric transmission along some sight-lines was affected by telluric ozone absorption lines. The wavelengths of the ozone features at $149.1558\, \mu$m and $149.7208\, \mu$m originating from the signal band coincide with the CH hyperfine transitions at $2010\,$GHz. These ozone lines, despite having relatively narrow spectral features, contaminate the CH absorption measurements made toward the strong continuum sources. Therefore, the receivers were tuned to 2006.7$\,$GHz in the upper side band mode. 

\begin{table*}
\caption{Spectroscopic parameters for $N,J= 2, 3/2 \rightarrow 1, 1/2$ hyperfine transitions of CH.}
\begin{center}

\begin{tabular}{ccccc}
\hline \hline
\multicolumn{2}{c}{Transition} &  Frequency & $A_\text{E}$  &  $E_{u}$  \\%&  $E_{l}$\\
Parity & $F$&  [GHz]   & $10^{-2} \times $ [s$^{-1}$]&     [K]      \\%&  [K]\\
\hline
 $  - \rightarrow + $& $1 \rightarrow 1$ & 2006.74892 & 1.117 & 96.31 \\%& 0.00072\\
                    & $1 \rightarrow 0$ & 2006.76263 & 2.234   \\%& 0.00000\\
					& $2 \rightarrow 1$ & 2006.79912 & 3.350 \\%& 0.00072\\
                    \hline
 $ + \rightarrow - \,\,^{*} $ & $1 \rightarrow 1$ & 2010.73859 & 1.128 & 96.66\\%& 0.16071\\
                    & $1 \rightarrow 0$ & 2010.81046 & 2.257 &  \\%& 0.15726\\
					& $2 \rightarrow 1$ & 2010.81192 & 3.385 &  \\%& 0.16071\\
\hline  
\label{tab:freq}
\end{tabular}
\end{center}
\tablefoot{Columns are quantum number information, frequency, Einstein A coefficient and upper level energy; all taken from the Cologne Database for Molecular Spectroscopy \citep{muller2005cologne}. The frequencies were determined by \citet{davidson2001measurement}. $^{(*)}$ - indicate the transitions that are not observed in this experiment.}
\end{table*}

\begin{figure*}
\includegraphics[width=0.5\textwidth]{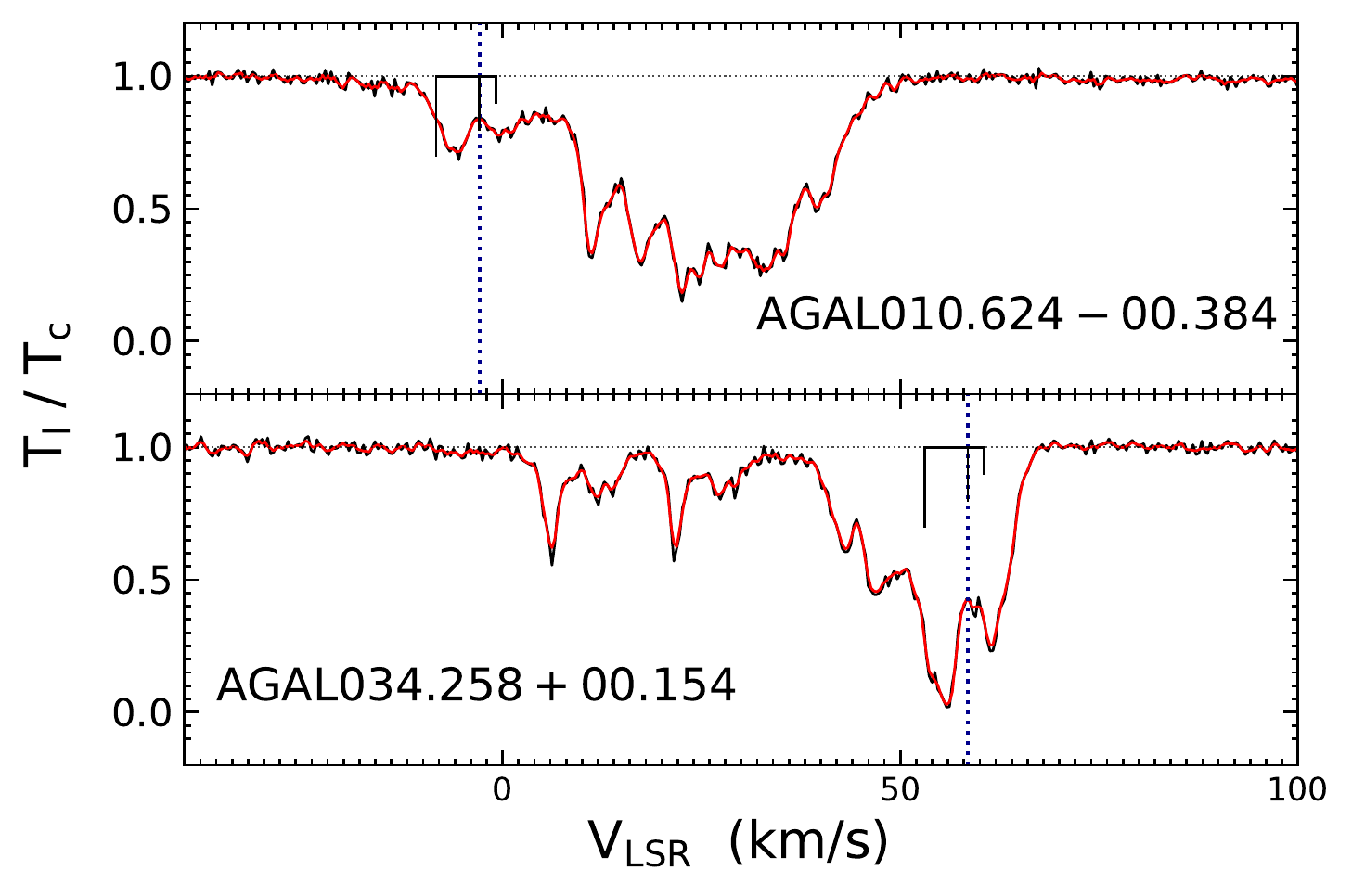}\quad
\includegraphics[width=0.49\textwidth]{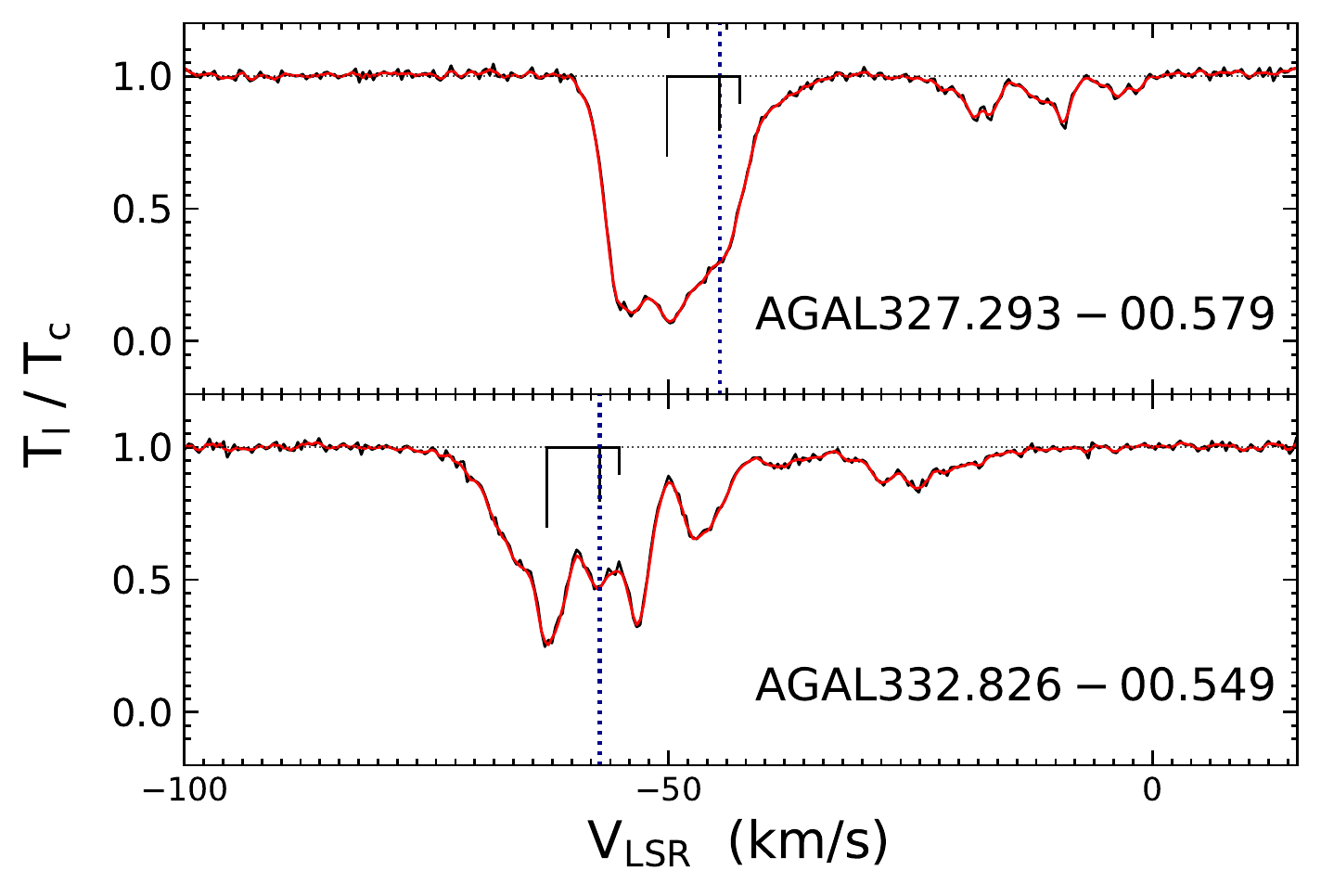}\\
\includegraphics[width=0.5\textwidth]{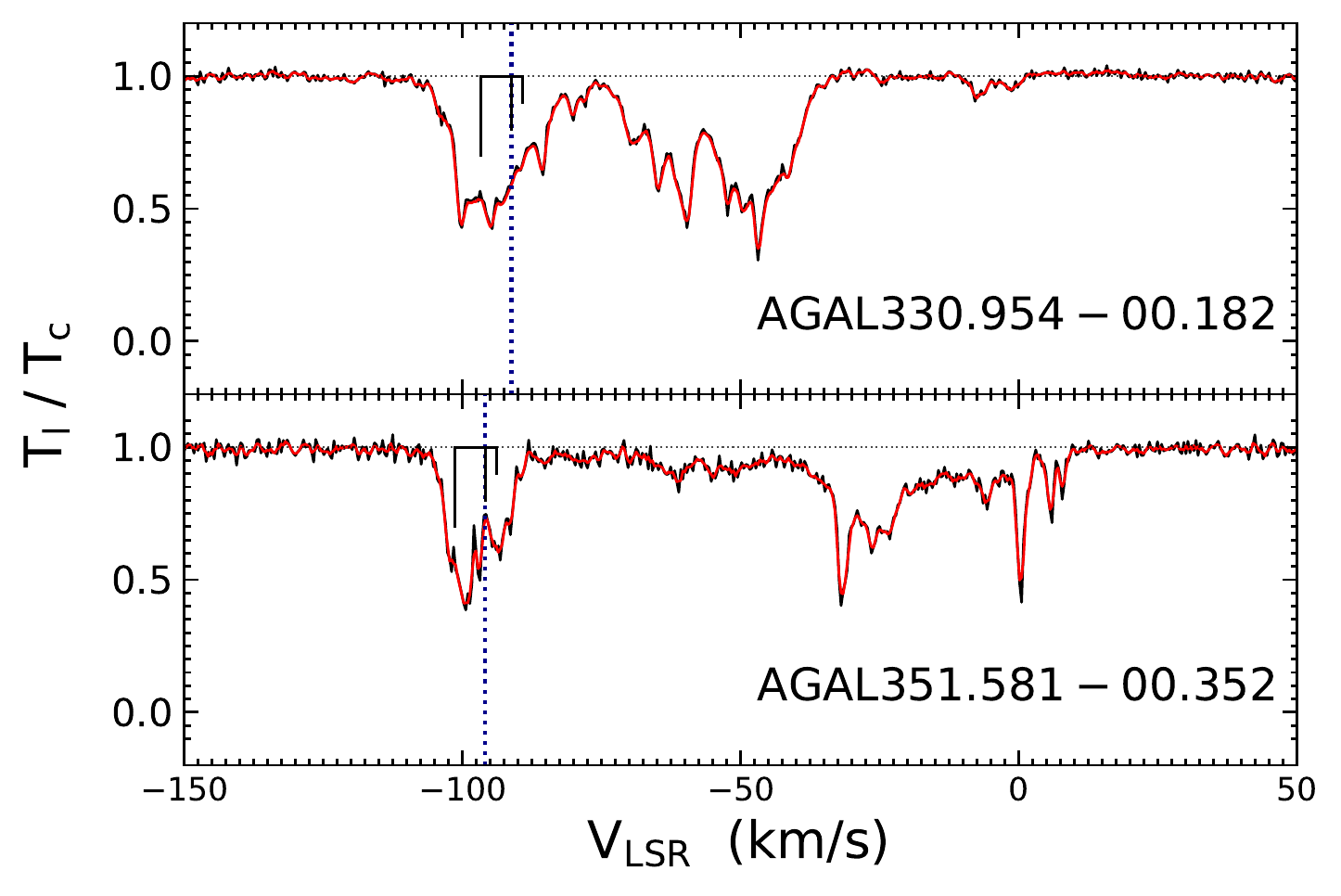}\quad
\includegraphics[width=0.505\textwidth]{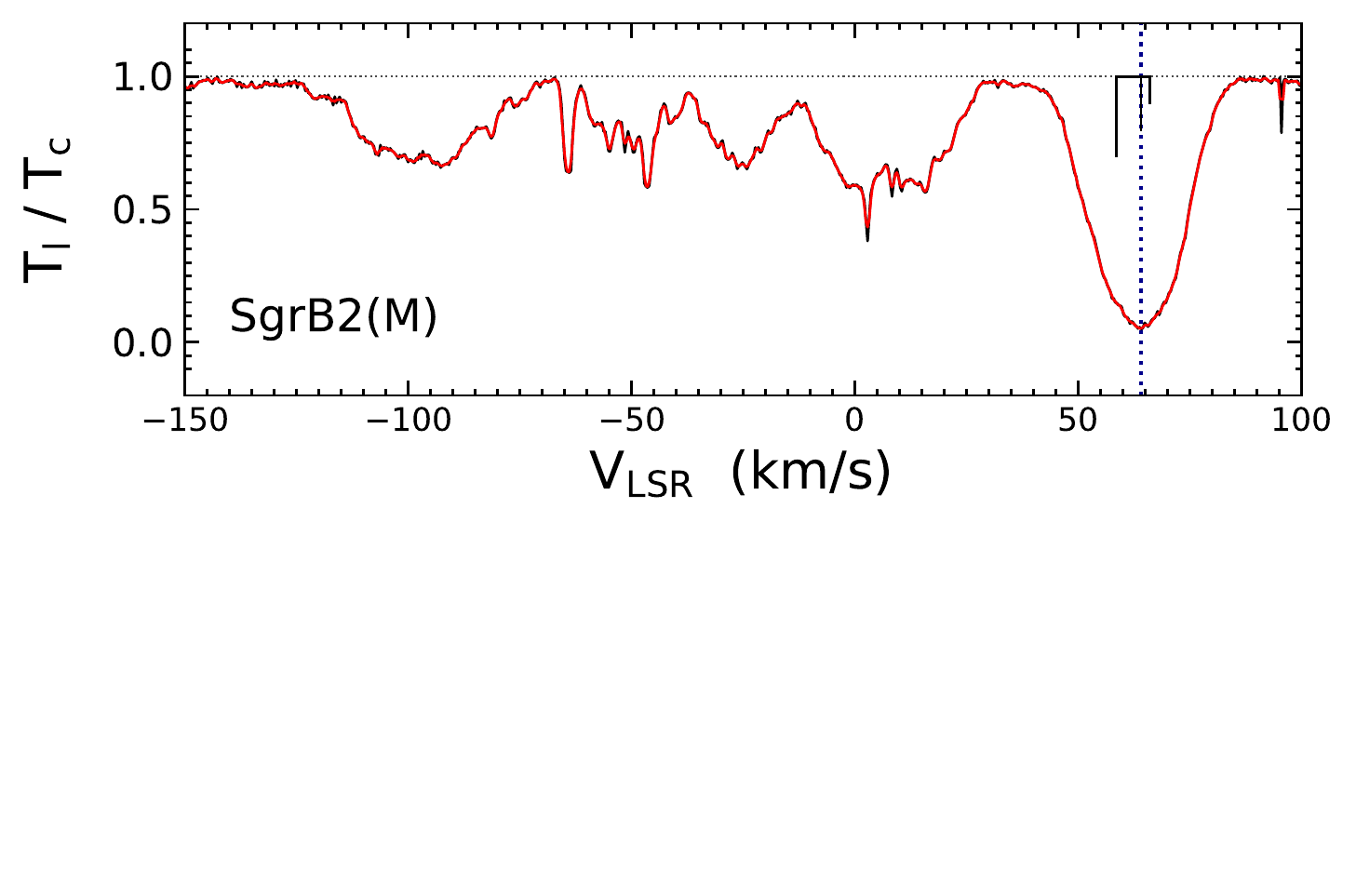}
\caption{CH ($N, J = 2, 3/2 \rightarrow 1, 1/2$) hyperfine transition spectra observed toward (from top) AGAL010.624$-00$.384, AGAL034.258+00.154, AGAL327.293$-00$.579, AGAL332.826$-00$.549, AGAL330.954$-00$.182, AGAL351.581$-00$.352, and SgrB2(M) using GREAT/SOFIA. Each spectrum is normalized with respect to its continuum level. The WF fits are overlaid in red and include the HFS structure. The positions and relative intensities of the HFS structure are displayed in black. Sources with similar velocity distributions were grouped to display the absorption features along the LOS, in detail. The systemic velocities of the sources are represented by the dotted blue lines.}
\label{fig:sofia_spec}
\end{figure*}

Both atmospheric as well as systematic fluctuations arising from the instrumental set-up were removed by using a double beam chop-nod mode. The secondary mirror chopped at a rate of $2.5\,$Hz with a chop amplitude of $60^{\prime \prime}$ and $105^{\prime \prime}$ for the observations carried out in 2017 and 2018. An advanced version of the MPIfR-built Fast Fourier transform spectrometer (dFFTS4G) \citep{dFFTS4G} was used as the backend to carry out the spectral analysis of the procured data. A velocity resolution of $0.036\,$km~s$^{-1}$ was achieved by using a $4\,$GHz bandwidth per pixel, with a channel spacing of $244\,$kHz over 16384 channels for almost all the $2006\,$GHz spectra. The spectra were further calibrated using the {\small KALIBRATE} program \citep{guan2012great} with an underlying forward efficiency of 0.97 and a main-beam efficiency of 0.68. The fully calibrated spectra were subsequently analysed using the GILDAS-CLASS software\footnote{Software package developed by IRAM, see \url{https://www.iram.fr/IRAMFR/GILDAS/} for more information regarding GILDAS packages.}. The spectra were smoothed to 0.36~km~s$^{-1}$-wide velocity bins, and up to a second order polynomial baseline was removed. After baseline subtraction, the continuum level determined by using a double-sideband (DSB) calibration, was added back to the spectra to obtain the correct line-to-continuum ratio. Based on the instrumental performance and atmospheric transmission, we assume a 5$\%$ error in the calibration of the continuum level. The relative sideband gain is calibrated by assuming a signal-to-image band gain ratio of unity, as determined by \citet{kester2017derivation} for all bands of the HIFI instrument, with an accuracy between 1 and 4\%. Hence, even for the upGREAT instrument sharing the same receiver technology as that of HIFI bands 6 and 7, we assume that there is no significant departure in the band gain ratio from unity.

The CH spectra are detected in deep, yet unsaturated absorption toward each individual sightline, even at velocities corresponding to star forming regions (SFRs) as shown in Fig.~\ref{fig:sofia_spec}. This is because almost all of the CH molecules are expected to remain in the ground state within dilute envelopes and diffuse clouds, surrounding the denser hot-core regions. The observed absorption features correspond to the radial velocities expected for different spiral arms, and are broadened by the kinematic structure of the intervening absorbing medium.

\begin{figure}
\includegraphics[width=0.51\textwidth]{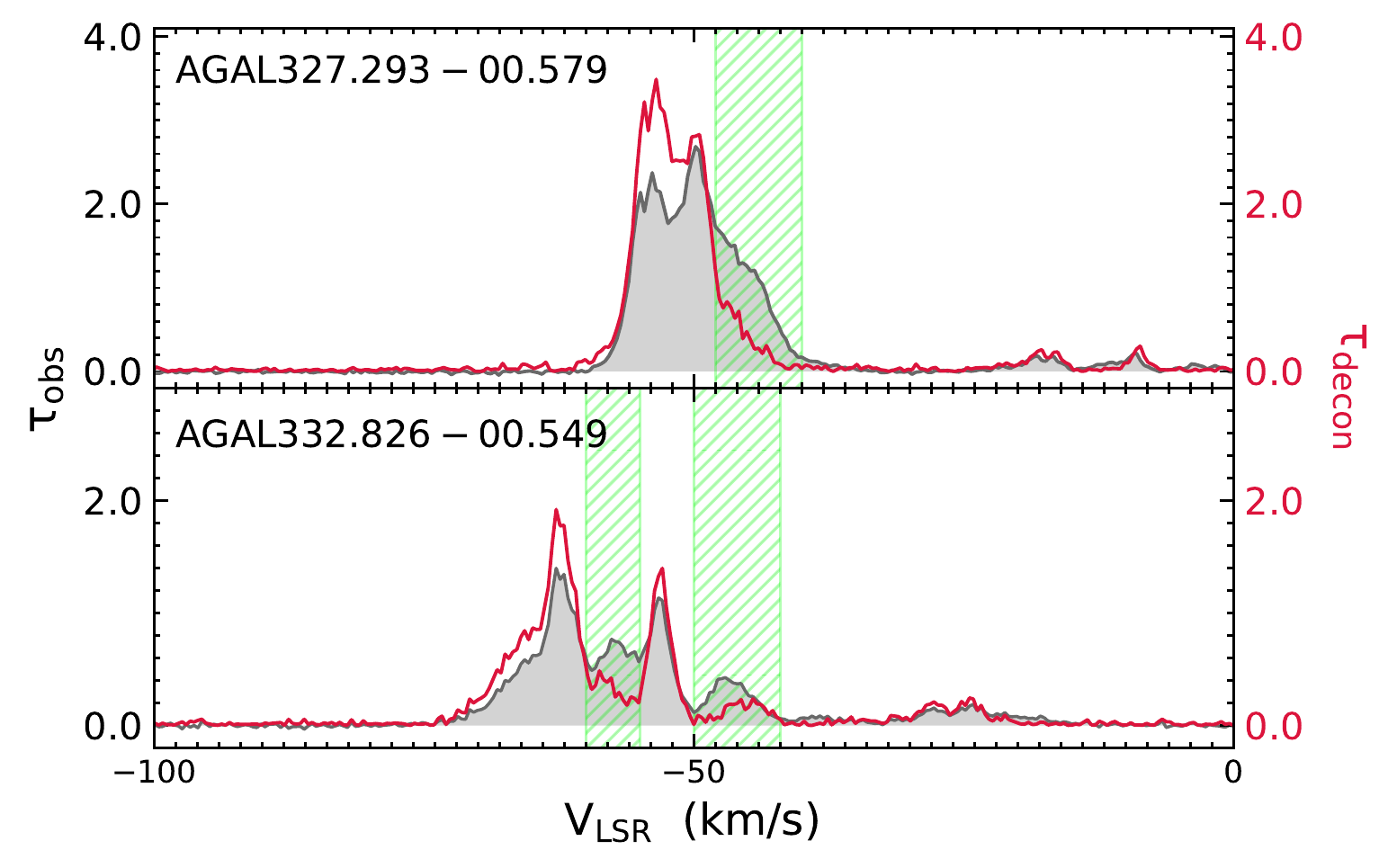}
\caption{Example illustrating effects of HFS splitting. The grey-filled area represents the observed spectra toward AGAL327.293$-00$.579 (top) and AGAL332.826$-00$.549 (bottom) in terms of optical depth, while their WF deconvolved spectra are displayed in red. The green hashed regions indicate velocity intervals where we observe line broadening due to HFS splitting. }
\label{fig:sofia_wf_decon}
\end{figure}

\section{Line profile analysis}\label{sec:analysis}

\subsection{WF deconvolution}\label{subsec:WFD_lineprofile}
In Fig.~\ref{fig:sofia_spec}, we display how the WF fits to the CH spectra observed towards the different lines of sight (LOS), following the algorithm described in Sect.~\ref{sec:WF_algo}. In Appendix~\ref{appendix:convolution_model}, we illustrate how the observed degraded spectrum is the convolution product of the deconvolved spectrum and the hyperfine weights. For two sources, we compare the line profiles of the observed spectrum with those of the deconvolved one in Fig.~\ref{fig:sofia_wf_decon}. The column densities determined from the former will be systematically larger over velocity intervals corresponding to the molecular clouds due to the line broadening effect of the HFS splitting, as evidenced from the differences between the two line profiles. While the extent of these variations are sightline-dependent, we find the column densities of CH determined from the observed spectra (prior to deconvolution) to be on average 24.8$\%$ larger than those determined post deconvolution, for the molecular cloud features. The quality of the WF fit toward the varied sightlines are assessed prior to deconvolution by means of the fit residuals (see Appendix~\ref{appendix:residuals}). Furthermore, we discuss the efficiency of the WF fit and deconvolution in Appendix~\ref{appendix:forward}.

The column densities are derived following Eq.~\ref{eq:coldens} by assuming an excitation temperature of $\sim3.1\,$K, which is close to the temperature of the cosmological blackbody radiation (2.738$\,$K), and was found to represent contributions of the Galactic interstellar radiation field \citep{gerin2010interstellar}, and integrated over velocity intervals thought to be characteristic of spiral-arm and inter-arm features, based on the spiral-arm structure of the Milky Way presented by \citet{reid2014trigonometric}, and consistent with those used by \citet{godard2012comparative} and \citet{wiesemeyer2016far}. The ground state occupation assumption at LTE ($T_\text{ex}$\,=\,$\,T_\text{rot}$\,=\,3.1\,K) is valid for the physical conditions that prevail over diffuse and translucent regions, and is questionable only within the molecular cloud components themselves. The dense envelopes of these molecular clouds can result in higher values of $T_\text{ex}$ as a consequence of collisional excitation of CH. The consequences of determining column densities by assuming a single excitation temperature component along each LOS is briefly discussed in Appendix~\ref{appendix:impact_excitationtemp}. Upon modelling the absorption features associated with the dense molecular environments of our continuum sources, we find the derived column densities of CH increase by $\sim15\%$ at $T_\text{ex}$\,=\,10~K, and up to $<50\%$ at $T_\text{ex}$\,=\,30~K. This further indicates that the uncertainties in our column density estimates at velocity intervals corresponding to those of the molecular cloud components predominantly arise from our assumption of a single excitation temperature. Hence, the column densities derived over their associated velocity components represent a lower limit on the $N$(CH) values. The derived column densities for each sightline are summarised in Table.~\ref{tab:coldens}. The CH column densities per velocity component are found to vary between $\sim 10^{12}$ and $1.2\times 10^{14}\,$cm$^{-2}$ along inter-arm and spiral-arm clouds along the different sightlines. We inspect the reliability of the column densities derived using the WF by comparing them with other techniques in Sect.~\ref{subsec:comparison}.
\begin{table*}
\caption{CH $N,J = 2,3/2 \rightarrow 1,1/2$ absorption line analysis using WF deconvolution algorithm and comparison of column densities with OH. }
\setlength{\tabcolsep}{10pt} % Default value: 6pt
\renewcommand{\arraystretch}{1.3} % Default value: 1
\label{tab:coldens}      
\begin{center}          
\begin{tabular}{ p{0.9cm} p{0.9cm}  p{0.9cm} p{0.9cm}  c c c c c }     %9 columns 
\hline\hline       
  $\varv_\text{min}$ & $\varv_\text{max}$ & $R_{g, \text{min}}$\tablefootmark{a} & $ R_{g, \text{max}}$\tablefootmark{a}  & Remark\tablefootmark{b} & $N$(CH) & $N$(\molh)\tablefootmark{c} & $N$(OH)\tablefootmark{d}%& $r$ 
  \\
     $[\text{km~s}^{-1}]$ & $[\text{km~s}^{-1}]$ & [kpc] & [kpc] & & {$[10^{13} \, \text{cm}^{-2}]$} & $[10^{21} \, \text{cm}^{-2}]$  & $[10^{14} \, \text{cm}^{-2}]$  \\ \hline
    \multicolumn{8}{c}{AGAL010.624$-00$.384 }\\
    \hline
     $-11$ & $\,\,-2$ & $\,\,\,$$8.9$ & 11.6 &  MC &$>4.82$ & -- &  $>3.42$ %&  0.14
    \\
        $\,\,-2$ & $\,\,+2$  & $\,\,\,$$8.1$ & $\,\,\,8.9$ &  MC   &$>4.84$ & -- & $>10.21$ %& 0.05
       \\
       $\,\,+2$ & $\,\,+7$ & $\,\,\,$$7.2$ & $\,\,\,8.1$  & Norma  & $\,\,\,$7.78$^{+0.22}_{-0.05}$& 2.22 & $\,\,\,$5.68$^{+0.17}_{-0.03}$ %& 0.14
       \\
        $\,\,+7$ & $+15$ & $\,\,\,$$6.2$ & $\,\,\,7.2$ & Sgr & 15.70$^{+0.05}_{-0.01}$ & 4.48 &  $\,\,\,$6.54$^{+0.09}_{-0.02}$ %& 0.24
       \\
 $+15$ & $+20$ & $\,\,\,$$5.7$ & $\,\,\,6.2$ &     & 25.26$^{+0.17}_{-0.08}$& 7.21 &  14.45$^{+0.10}_{-0.05}$ %& 0.17 
       \\
        $+20$ & $+31$ & $\,\,\,$$4.8$ & $\,\,\,5.7$ & Scu/Per    & 27.31$^{+0.06}_{-0.02}$&7.80 &  17.67$^{+0.14}_{-0.06}$ %&  0.15 
       \\
        $+31$ & $+36$ & $\,\,\,$$4.5$ & $\,\,\,4.8$ &   & 22.63$^{+0.10}_{-0.00}$& 6.46 & S %& --
       \\
        $+36$ & $+51$ & $\,\,\,$$3.7$ & $\,\,\,4.5$ & Sgr    & $\,\,\,$5.31$^{+0.19}_{-0.01}$ & 1.51 & S %& --
       \\ \hline

    \multicolumn{8}{c}{AGAL034.258+00.154 }\\
    \hline
  $\,\,\,+0$ & $+16$  & $\,\,\,$$7.5$ & $\,\,\,8.5$ &   & $\,\,\,$1.65$^{+0.28}_{-0.33}$& 0.47 & $2.71^{+0.11}_{-0.03}$%& 0.06
\\
        $+16$ & $+32$ & $\,\,\,$$6.8$ & $\,\,\,7.5$ & Sgr  & $\,\,\,$1.51$^{+0.20}_{-0.17}$ & 0.43 & $1.81^{+0.08}_{-0.11}$ %& 0.08 
       \\
        $+32$ & $+38$ & $\,\,\,$$6.5$ & $\,\,\,6.8$ &      & $\,\,\,$0.76$^{+1.37}_{-0.84}$ & 0.22 &$0.48^{+0.82}_{-0.61}$ %&  0.15
       \\
        $+38$ & $+68$ & $\,\,\,$$5.5$ & $\,\,\,6.5$ &   MC   & $>9.35$ & -- & S %& --
       \\ \hline 
       \multicolumn{8}{c}{AGAL327.293$-00$.579}\\
    \hline
 $-61$ & $-32$ & $\,\,\,$7.4 & $\,\,\,$9.5 & Scu, MC & $>9.68$ & -- & S %& --
\\
         $-32$ & $-25$ & $\,\,\,$9.5 & 10.2& & $\,\,\,$$0.22^{+0.38}_{-0.41}$ & 0.06 & $0.05^{+0.71}_{-0.38}$ %& 0.44
        \\
         $-25$ & $-14$ & 10.2 & 11.5 & & $\,\,\,$$1.10^{+0.50}_{-0.09}$ & 0.31 & $0.52^{+0.25}_{-0.15}$ %& 0.21 
        \\
         $-14$ & $ \,\,\,-6$ & 11.5 & 12.6 & & $\,\,\,$$0.82^{+1.10}_{-0.05}$ & 0.23 & $0.98^{+0.24}_{-0.18}$ %& 0.08
        \\
         $\,\,\,-6$ & $\,\,\,+3$ & 12.6 & 14.2 &  & $\,\,\,$$0.22^{+0.85}_{-1.08}$ & 0.06 & $0.42^{+0.30}_{ -0.15}$ %& 0.20 
        \\ \hline
         \multicolumn{8}{c}{AGAL330.954$-00$.182}\\
    \hline
  \hspace{-0.195cm}$-115$ & $-75$ & $\,\,\,$4.1 & $\,\,\,$5.0 & MC & $>2.96$ & -- & S %& -- 
\\
         $-75$ & $-55$ & $\,\,\,$5.0 & $\,\,\,$5.6 &  & $\,\,\,$$3.80^{+0.05}_{-0.12}$& 1.09 & 1.23$^{+0.05}_{-0.02}$ %& 0.31
        \\
         $-55$ & $-30$ & $\,\,\,$5.6 & $\,\,\,$6.6 & Norma & $\,\,\,$$3.66^{+0.01}_{-0.02}$ & 1.04 & S %& --
        \\
         $-30$ & $-10$ & $\,\,\,$6.6 & $\,\,\,$7.7 & Scu & $\,\,\,$$0.19^{+1.31}_{-1.07}$ & 0.05 &  $0.04^{+0.70}_{-0.51}$ %& 0.53 
        \\
         $-10$ & $\,\,\,+4$ & $\,\,\,$7.7 & $\,\,\,$8.8 & Norma & $\,\,\,$$0.45^{+0.36}_{-0.31}$ & 0.12 & $0.09^{+0.03}_{-0.02}$ %& 0.50 
        \\ \hline
        \multicolumn{8}{c}{AGAL332.826$-00$.549 }\\
    \hline
    $-81$ & $-50$ & $\,\,\,$4.7 & $\,\,\,$5.6 & Norma, MC & $>5.00$& -- & S %& --
\\
         $-50$ & $-40$ & $\,\,\,$5.6 & $\,\,\,$6.0 & & $\,\,\,$$1.12^{+0.02}_{-0.02}$& 0.32 & $5.20^{+0.09}_{-0.02}$ %& 0.02
        \\
         $-40$ & $-32$ & $\,\,\,$6.0 & $\,\,\,$6.4 &  & $\,\,\,$$0.34^{+0.60}_{-0.52}$ & 0.09 & $3.94^{+0.13}_{-0.01}$ %& 0.008
        \\
         $-32$ & $\,\,\,$$-8$ & $\,\,\,$6.4 & $\,\,\,$7.8 & Sgr & $\,\,\,$$0.72^{+0.75}_{-0.67}$ & 0.20 & $2.57^{+0.11}_{-0.01}$ %& 0.03 
        \\ \hline
        
        \end{tabular}
\end{center}
\end{table*}
\addtocounter{table}{-1}
\begin{table*}
\caption{Continued.}
\setlength{\tabcolsep}{10pt} % Default value: 6pt
\renewcommand{\arraystretch}{1.3} % Default value: 1
\label{tab:coldens1}      
\begin{center}          
\begin{tabular}{ p{0.9cm} p{0.9cm}  p{0.9cm} p{0.9cm}  c c c c c }     %9 columns 
\hline\hline       
  $\varv_\text{min}$ & $\varv_\text{max}$ & $R_{g, \text{min}}$\tablefootmark{a} & $ R_{g, \text{max}}$\tablefootmark{a}  & Remark\tablefootmark{b} & $N$(CH) & $N$(\molh)\tablefootmark{c} & $N$(OH)\tablefootmark{d}%& $r$ 
  \\
     $[\text{km~s}^{-1}]$ & $[\text{km~s}^{-1}]$ & [kpc] & [kpc] & & {$[10^{13} \, \text{cm}^{-2}]$} & $[10^{21} \, \text{cm}^{-2}]$  & $[10^{14} \, \text{cm}^{-2}]$  \\ \hline
        
        \multicolumn{8}{c}{AGAL351.581$-00$.352}\\
    \hline
  \hspace{-0.195cm}$-109$ & $-88$ & 2.0 & $\,\,\,$2.3 & MC & $>3.65$ & -- & S %& --
\\
         $-88$ & $-68$ & 2.3 & $\,\,\,$2.7  & & $\,\,\,$$0.45^{+0.30}_{-0.25}$ & 0.13 & $0.20^{+0.10}_{-0.03}$ %& 0.23 
        \\
         $-68$ & $-42$ & 2.7 & $\,\,\,$3.7 & Sgr & $\,\,\,$$0.81^{+0.31}_{-0.23}$ & 0.23 & $0.94^{+0.05}_{-0.01}$ %& 0.08
        \\
         $-42$ & $-14$ & 3.7 & $\,\,\,$6.0 & Norma & $\,\,\,$$2.69^{+0.06}_{-0.04}$ & 0.76 & S %& -- 
        \\
         $-14$ & $-12$ & 6.0 & $\,\,\,$6.2 & & $\,\,\,$$0.90^{+0.24}_{-0.20}$ & 0.25 & $0.15^{+1.06}_{-0.08}$ %& 0.60
        \\
         $-12$ & $\,\,\,$$-3$ & 6.2& $\,\,\,$7.7& Sgr& $\,\,\,$$1.80^{+0.11}_{-0.07}$ & 5.14 & $0.18^{+0.26}_{-0.08}$ %& 0.96
        \\
         $\,\,\,$$-3$ & $\,\,\,$$+3$ & 7.7 & $\,\,\,$9.4 & Scu & $\,\,\,$$2.99^{+0.06}_{-0.04}$ & 0.85 & S %& -- 
        \\
         $\,\,\,$$+3$ & $+12$ & 9.4 & 13.5 & & $\,\,\,$$0.44^{+0.40}_{-0.24}$ & 0.12  & $0.28^{+0.02}_{-0.01}$ %& 0.16
        \\ \hline 
\multicolumn{8}{c}{SgrB2(M)}\\
\hline
\hspace{-0.195cm}$-124$ & $-67$ & & & & $27.81^{+0.01}_{-0.02}$ & 7.94 & -- %& --   
\\
          $-67$ & $-38$ & & & & $\,\,\,$$3.16^{+0.10}_{-0.13}$ & 0.90 & -- 
         %& -- 
         \\
          $-38$ & $-12$ & & & 3~kpc & $\,\,\,$$3.60^{+0.09}_{-0.08}$&1.30& -- %& --
         \\
          $-12$ & $\,\,\,$$+8$ & & & GC & $\,\,\,$$6.28^{+0.01}_{-0.02}$ & 1.79 & -- %& --
         \\
          $\,\,\,$$+8$ & $+29$ & & & Sgr & $\,\,\,$$4.57^{+0.11}_{-0.11}$ & 1.31 & -- %& --
         \\
          $+29$ & $+40$ & & & Scu & $\,\,\,$$0.89^{+0.61}_{-0.49}$ & 0.25 & -- %& --
         \\
          $+40$ & $+90$ & & &MC  & $>73.02$ &-- & -- %& -- 
         \\ \hline
\end{tabular}
\end{center}
\tablefoot{\tablefoottext{a}{The galactocentric distances for each velocity interval is computed using, $R_\text{G} = R_{0}\frac{\Theta(R_\text{G})\text{sin}(l)\text{cos}(b)}{v_\text{lsr} + \Theta_{0}\text{sin}(l)\text{cos}(b)}$, with $R_{0}$ = 8.3$\,$kpc, $\Theta_{0}$ = 240$\,$km~s$^{-1}$ and assuming a flat Galactic rotation curve, i.e, $\Theta(R_\text{G}) = \Theta_0$. Note that we do not account for the non-circular motion of the 3~kpc arm \citep{dame2008new}.}\tablefoottext{b}{The positions of the four main spiral arms and the local arms have been inferred from the spiral arm characteristics presented in \citet{reid2014trigonometric} and the velocities corresponding to the molecular cloud envelopes are indicated by MC.} \tablefoottext{c}{$N(\text{H}_{2})$ is derived from [CH]/[\molh] = $3.5\times 10^{-8}$ given by \citet{sheffer2008ultraviolet}}. \tablefoottext{d}{S denotes velocity components with saturated absorption line profiles.}}
\end{table*}

As the optical depths are computed from the line-to-continuum ratio, uncertainties in the continuum level give rise to systematic errors in the derived column densities. The uncertainties are partly due to the receiver-gain ratio between the signal and the image band, possibly deviating from unity. For the mixer technology (hot-electron bolometers) used by upGREAT, this quantity is difficult to measure without re-tuning the receiver. For the HIFI bands employing the same technology and closest to our tuning, a dedicated study \citep{kester2017derivation} provides typical deviations of 1 to 4\%. At this level, given the available sensitivity and typical optical depths ranging from 0.1 to 1, the impact of the continuum uncertainty on the derived column densities is difficult to disentangle from that of the thermal noise in the absorption profile. However, following the DSB error estimation presented in \cite{wiesemeyer2018unveiling}, we account for the uncertainties present in the continuum level that will subsequently lead to errors in the optical depths and derived column densities. Further, the errors in the WF computed column densities are determined by sampling a posterior distribution of the deconvolved optical depths\footnote{The error in the derived column densities (per velocity interval) scale with the deconvolved optical depths by a constant value. This constant is a function of the spectroscopic parameters that govern the transition and the excitation temperature.}. We sampled 5000 artificial spectra, each generated by adding a pseudo-random noise contribution with the same standard deviation as the line free part of the continuum, to the absorption spectra prior to applying the WF fit and deconvolution, over each iteration. The deconvolved optical depths and subsequently derived column densities (per velocity interval) sample a point in the channel-wise distribution of the column densities across all spectra. The empirical spread of these distributions yield the asymmetric errors in the computed column densities. However, the column densities are not always normally distributed, and often showcase skewed distributions, as normality is not imposed. For this reason, the profiles of the distributions are best described by using the median and inter-quartile range from which the sample mean and standard deviation are derived following \citet{wan2014estimating}. This is done so as to remove any biases introduced in the determination of the mean due to the asymmetry of the distribution. The new mean and standard deviation are then used to determine the asymmetric errors through positive and negative deviations. While errors in the deep absorption features corresponding to spiral-arm velocities arise from imperfections in the fit, the errors in the inter-arm gas features are dominated by uncertainties in the true continuum.

\subsection{Comparison of WF algorithm with other procedures}\label{subsec:comparison}

In this section, we briefly compare column densities of the CH molecule inferred using the WF algorithm with those obtained from other commonly used techniques. To carry out an unbiased analysis, we performed the WF algorithm on the $N=1, J = 3/2 \rightarrow 1/2$ HFS spectra of CH near 532$\,$GHz, obtained using \textit{Herschel}/HIFI toward SgrB2(M) as a part of the HEXOS\footnote{The guaranteed time key project, \textit{Herschel}/HIFI observations of EXtra-Ordinary Sources(HEXOS) \citep{bergin2010herschel} was aimed to investigate the chemical composition of several sources in the Orion and SgrB2 star-forming regions.} survey toward SgrB2(M). The spectroscopic parameters of these HFS transitions are summarised in Table.~\ref{tab:freqothers}. The LOS toward SgrB2(M) was chosen because of its complex spectral line profile which contains an amalgamation of narrow and broad features, alike. \citet{qin2010herschel} modelled the CH spectra by using XCLASS \citep{moller2017extended} and the automated fitting routine provided by MAGIX\footnote{See \url{https://magix.astro.uni-koeln.de/} for more information about the MAGIX software.} for an excitation temperature of 2.73$\,$K. Assuming LTE, XCLASS solves the radiative transfer equation with an underlying Gaussian brightness profile to model the observed spectral line profile. Using the WF algorithm, we derived the column densities of the CH HFS transitions near 532$\,$GHz, integrated over the same velocity intervals as those specified in Table.~1 of \citet{qin2010herschel}. The errors were computed as before, but with the assumption of a $20\%$ error in the continuum level.
\begin{figure*}
    \sidecaption
    \includegraphics[width=12cm]{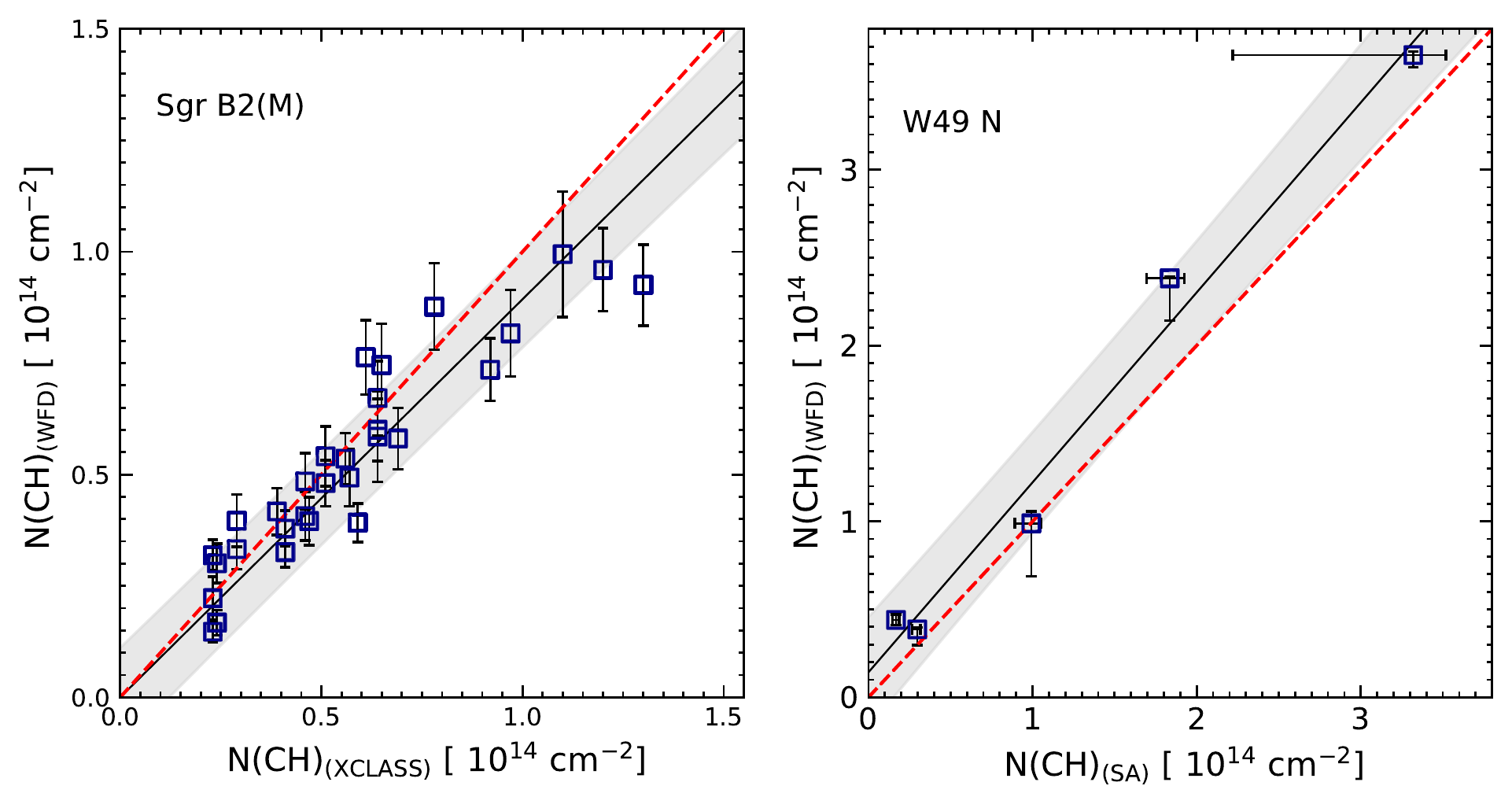}
    \caption{Left: Comparison of $N=1, J=3/2 \rightarrow 1/2$ CH column densities derived using WF deconvolution and the XCLASS software presented in \citet{qin2010herschel}. Right: Comparison of $N,J= 2, 3/2 \rightarrow 1, 1/2$ CH column densities derived using WF deconvolution and the simulated annealing algorithm presented in \citet{wiesemeyer2018unveiling}. The black solid line represents the weighted fit to the data, while the red dashed line represents a ratio of one. The data points that lie within $1\sigma$ intervals of the weighted linear regression are indicated by the grey-shaded regions. }
    \label{fig:compare}
\end{figure*}

A comparison between the CH column densities derived using the WF algorithm and the XCLASS fits is displayed in Fig.~\ref{fig:compare} for all absorption features along the LOS except, for the strong absorption from the envelope of the SgrB2(M) molecular cloud, which traces denser gas. The weighted regression at lower column densities systematically follows a one-to-one correlation while at larger column densities the values derived using the WF algorithm are almost always smaller than those computed using XCLASS, with an overall ratio of \textit{N}(CH)$_\text{WFD}$/\textit{N}(CH)$_\text{XCLASS}$ = $0.76 \pm 0.06$. The deviations may arise due to the intrinsically different means by which the two methods account for HFS splitting, an effect that is particularly prominent at the larger column density values. Additionally, inconsistencies in the post-processing of the \textit{Herschel}/HIFI data (baseline fitting, etc) also contribute to the observed deviations. Nonetheless, the CH abundances determined using both methods are in agreement within the systematic uncertainties. 
 
In addition, we also compare the column densities obtained using the WF deconvolution with those obtained by \citet{wiesemeyer2018unveiling} when using simulated annealing (SA) to the $N,J= 2, 3/2 \rightarrow 1, 1/2$ hyperfine transitions of CH toward W49~N. The column densities derived using both methods have a tight relation with a slope of, \textit{N}(CH)$_\text{WFD}$/\textit{N}(CH)$_\text{SA}$ = $1.08 \pm 0.08$. The column densities derived using the two methods deviate in value only within the inter-arm gas components and the overall concordances between the two methods further validates the use of the non-iterative WF deconvolution scheme.

\section{Results} \label{sec:results}
In Sect.~\ref{subsec:choh_col_dens}, we investigate the correlation between the column densities of CH with values determined for another widely studied light hydride, hydroxyl (OH), and evaluate its abundance using CH as a proxy for H$_{2}$. To further appreciate the global characteristics of the CH molecule as a surrogate for \molh within the Milky Way, in Sect.~\ref{subsec:radial_dist}, we study its radial distribution across the Galactic plane.

\subsection{CH versus OH column density}\label{subsec:choh_col_dens}

 For quite some time, CH has been used as a tracer for \molh in the diffuse regions of the ISM based on its observed correlation with \molh \citep{federman1982diffuse, mattila1986radio, sheffer2008ultraviolet}. As a precursor to the formation of CO, OH has also gained recognition as a promising tracer for the ``CO-dark" component \citep{grenier2005unveiling} of molecular gas in the ISM \citep{tang2017oh, li2018oh, engelke2018oh, rugel2018oh}. In Table~\ref{tab:coldens}, we present column densities of \molh estimated using the $N$(CH)/$N$(H$_{2}$)$\sim 3.5^{+2.1}_{-1.4} \times 10^{-8}$ relation derived by \citet{sheffer2008ultraviolet}. This correlation was established using optical observations of the $\text{A}^2\Delta$ - $\text{X}^2\Pi$ system of CH at $4300\,$\AA{} and the (2-0), (3-0), and (4-0) bands of the Lyman B-X transitions of \molh toward 37 bright stars. Akin to the CH observations presented in our study, the optical CH and UV \molh lines presented by \citet{sheffer2008ultraviolet} are seen in absorption toward sightlines that probe molecular clouds in the local diffuse ISM ($10^{19} < N(\text{H}_{2})<10^{21}\,$cm$^{-2}$).

\begin{figure}
\vspace{0.8cm}
    \centering
    \includegraphics[width=0.5\textwidth]{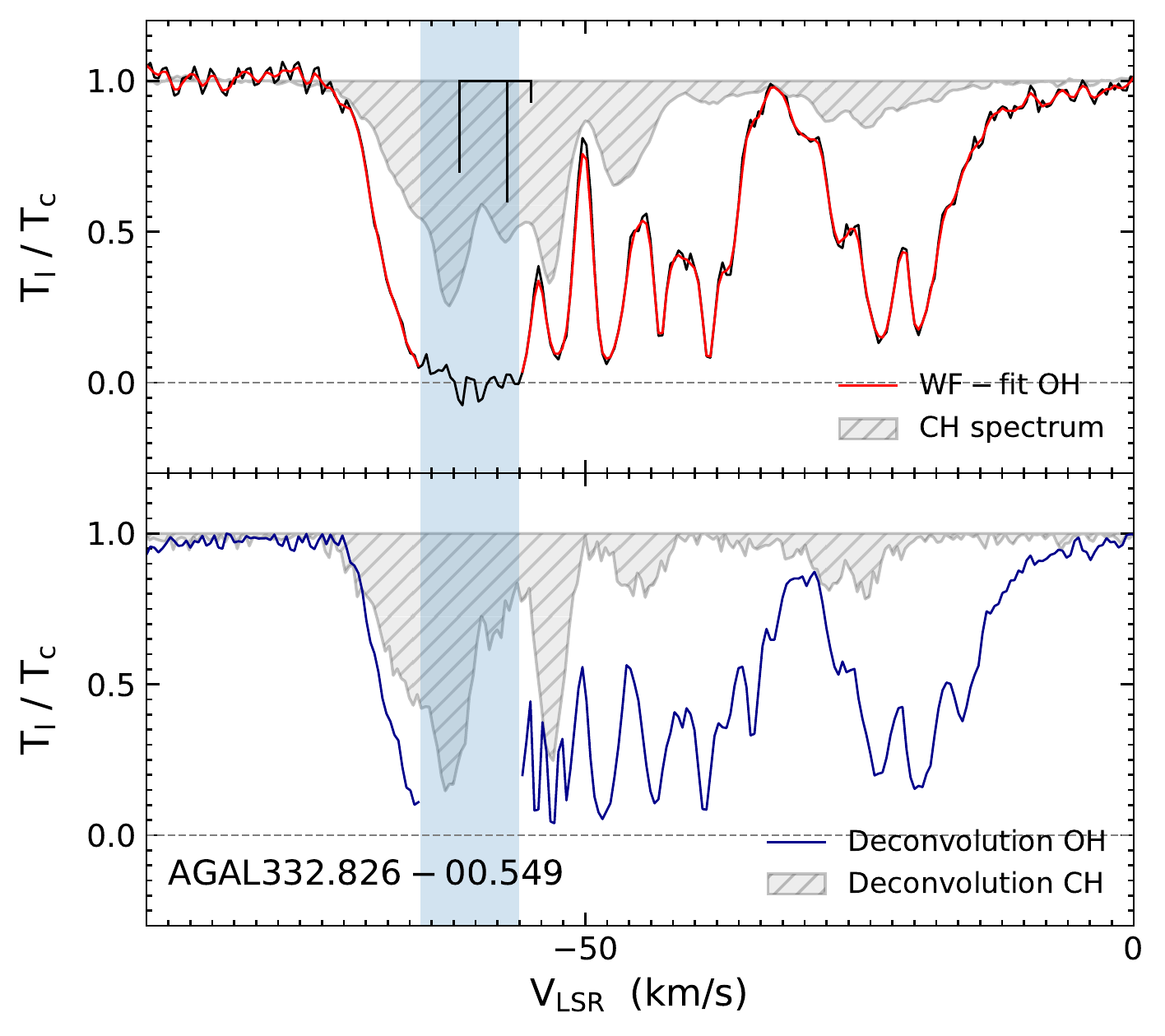}
    \caption{Top: Normalised line profile and HFS structure of OH near 2514~GHz (in black) overlaid with WF fit in red toward AGAL332.826$-00$.549. The WF fit avoids velocity intervals of the OH spectrum that show saturation between -65 < $\varv_\text{lsr}$ <-56$\,$km~s$^{-1}$, bounded by the blue-shaded regions. The grey shaded and hatched region displays the CH absorption profile toward the same source for comparison. Bottom: Normalised WF deconvolved line profile of OH (avoiding the saturated region) in blue, and corresponding deconvolved spectrum of CH displayed by the shaded and hatched grey region.   }
    \label{fig:chvsoh}
\end{figure}

 In order to carry out a comparison between $N$(OH) and $N$(CH), we use data published in \citet{wiesemeyer2016far} of the ${}^{2}\Pi_{3/2},\,J = 5/2 \rightarrow 3/2$ ground-state transitions of OH, procured using GREAT/SOFIA as a part of the observatory's cycle 1 campaign. The spectroscopic parameters of these OH transitions are summarised in Table.~\ref{tab:freqothers}. For this preliminary search, observations were carried out toward all the sightlines discussed in this paper, except for SgrB2(M). The OH spectra are saturated at velocities corresponding to the SFR-related molecular cloud components (that provide the background continuum emission) themselves, as shown in Fig.~\ref{fig:chvsoh}. The saturated absorption features of the OH spectra arising in the envelopes of the hot cores of these star-forming regions are indicative of the fact that a complete ground-state occupation assumption is no longer valid \citep{csengeri2012sofia}. Moreover, from Fig.~\ref{fig:chvsoh}, it can be seen that most of the observed OH absorption features have corresponding features in the CH spectrum with notable differences at velocity intervals between $-$13 and $-$21~km~s$^{-1}$, and from $-$36 to $-$42~km~s$^{-1}$, in this example. The WF fit and deconvolution is applied to the OH spectra with a $T_\text{ex} = 3.1\,$K (for consistency), and integrated over the same velocity intervals as those used for CH. Our assumption of a complete ground-state occupation for OH is broken down for spectral features associated with the dense molecular clouds. The WF faces singularities when $T_{\text{l}}$ tends to zero and thus avoids the OH features arising from these regions because they showcase saturated absorption. Therefore, the use of a lower excitation temperature does not affect our analysis.
 
\begin{comment}
 \begin{table}
\caption{Spectroscopic parameters for the $N,J= 2, 5/2 \rightarrow 1, 3/2$ hyperfine transitions of OH.}
\begin{center}
\begin{tabular}{ccccc}
\hline \hline
\multicolumn{2}{c}{Transition} &  Frequency & $A_\text{E}$  &  $E_{u}$ \\%&  $E_{l}$\\
Parity & $F^{\prime} \rightarrow F^{\prime \prime} $&  [GHz]   & $10^{-1} \times $ [s$^{-1}$]&     [K]     \\%&  [K]\\
\hline
 $ + \rightarrow - $& $2 \rightarrow 2$ & 2514.29805 & 0.1368 & 120.75065 \\%& 0.00072\\
                    & $3 \rightarrow 2$ & 2514.31640 & 1.3678 & 120.75153 \\%& 0.00000\\
					& $2 \rightarrow 1$ & 2514.35318 & 1.2310 & 120.75071 \\%& 0.00072\\
                    \hline
\label{tab:freqoh}
\end{tabular}
\end{center}
%\tablefoot{Columns (3),(4) and (5): Hyperfine frequencies, Einstein A-coefficient and upper level energies. The spectroscopic data were taken from the CDMS database ~\citep{muller2001cologne}.  }
\end{table}
\end{comment}

\begin{figure}
    \centering
    \includegraphics[scale = 0.6]{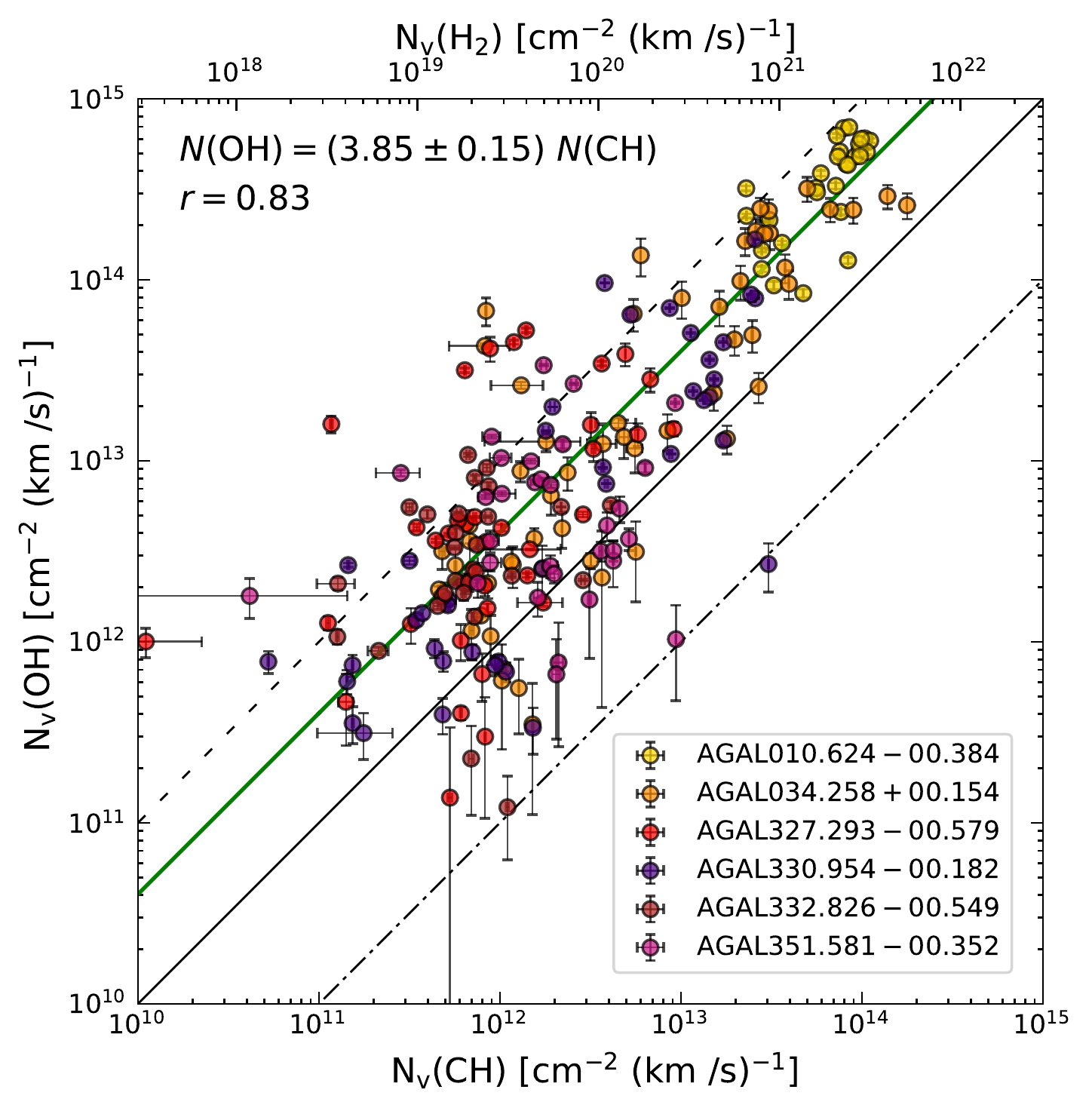}
    \caption{$N$(OH) vs. $N$(CH), per unit velocity interval derived using WF algorithm. The different coloured data points represent contributions from the different sources. The black dashed, solid, and dotted-dashed lines indicate $N$(OH)/$N$(CH) ratios of 10, one, and 0.1, respectively while the linear regression fit to the data, $N(\text{OH}) = (3.85 \pm 0.15)~N(\text{CH})$ is displayed in green.}
    \label{fig:scatter_nch_noh}
\end{figure}

The derived CH and OH column densities are presented in Fig.~\ref{fig:scatter_nch_noh}. The used data correspond to column densities based on values computed within the velocity intervals used in Table.~\ref{tab:coldens}. The plotted data points represent values re-sampled to 1~km~s$^{-1}$ wide velocity bins (except towards the velocity intervals corresponding to the molecular clouds, as well as those contaminated by outflows). The $N$(CH)-$N$(OH) correlation plots for the individual sources are presented in Appendix~\ref{appendix:correlations}. Across the different sightlines, we obtained a range of $N$(OH)/$N$(CH) values between one and 10. Moreover, almost all the sources, except AGAL351.581$-00$.352 and AGAL332.826$-00$.549, have significant correlation coefficients between 0.46 and 0.75 with false-alarm probabilities below 5$\%$ (summarised in Table.~\ref{tab:individual_correlations}) within the spiral arm data, while there is no remarkable correlation observed in the inter-arm regions. The lack of, or weak correlations present in the above mentioned sources can be attributed to the fact that the OH spectra toward these two sources showcase more saturated features along the LOS than in the remainder of the sources.

\begin{figure}
    \centering
    \includegraphics[scale = 0.6]{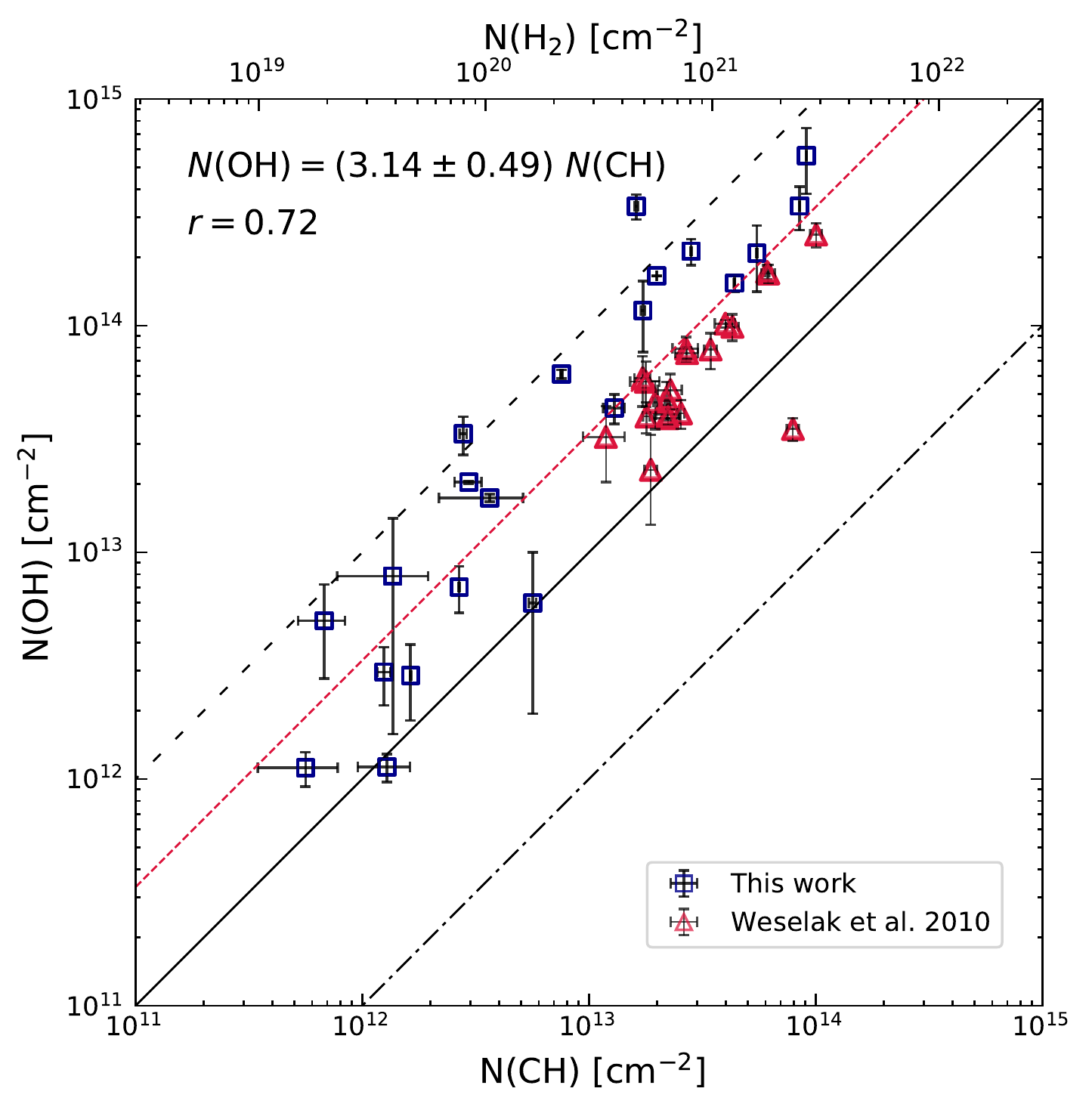}
    \caption{Correlation plot between column densities of the CH, and OH molecules, integrated over broad velocity intervals as per Table.~\ref{tab:coldens}. The red triangles and blue squares represent column density measures from \citet{weselak2010relation} and this work, respectively. The black dashed, solid, and dotted-dashed lines indicate $N$(OH)/$N$(CH) ratios of 10, one, and 0.1, respectively while the linear regression fit to the data, $N(\text{OH}) = (3.15 \pm 0.49)~N(\text{CH})$ is displayed by the red dotted line.}
    \label{fig:scatter_nch_noh_weselak}
\end{figure}

Fitting a linear regression to the combination of all the sightlines yields $N_\text{v}$(OH)/$N_\text{v}$(CH) = $(3.85 \pm 0.15)$, with a Pearson's correlation coefficient of 0.83 (false-alarm probability below $1\%$) for velocity intervals of 1~km~s$^{-1}$. 
The functional form of the regression was chosen after comparing the standard error of a regression fit through the origin with that of an ordinary least square regression, both of which yielded best fit slopes that are consistent with one another (within a $5\%$ error). From Fig.~\ref{fig:scatter_nch_noh}, we find that this relation spans roughly two orders of magnitude between 0.1 and 10, with a majority of the ratios lying between one and 10. We noted that those data points with $N$(OH)/$N$(CH)$\, > 10$ correspond to the broad line wings of saturated OH features, while those with $N$(OH)/$N$(CH)$\, < 1$ have larger errors in the determined $N$(OH) arising from their continuum-level uncertainties. Since the spectral features arising from star-forming regions are unsaturated in the case of the CH molecule, its effective broadening into the inter-arm regions is less. We further extend this analysis by combining our data set with optical observations\footnote{These observations were made using the UVES spectrograph at the European Southern Observatory (ESO) Paranal in Chile.} of the CH B-X transitions at 3886 and 3890$\, \text{\AA}$, and OH A-X transitions at 3078 and 3082$\, \text{\AA}$ presented toward 20 bright OB-type stars at low Galactic latitudes ($-17^{\circ}< b < +23^{\circ}$) from \citet{weselak2010relation}. Moreover, \citet{weselak2010relation} show that the LOS toward these stars probe physical conditions (gas densities, $n$(H) = 250-600$\,$cm$^{-3}$ and temperatures, $T_\text{kin} = 20-30\,$K) that are typical for diffuse molecular clouds \citep{snow2006diffuse} similar to those of our data set, and report a $N$(OH)/$N$(CH) value of 2.52$\pm$0.35. The revised abundance ratio (combining both sets of data) displayed in Fig.~\ref{fig:scatter_nch_noh_weselak} follows $N$(OH)/$N$(CH) = $(3.14 \pm 0.49)$. Using this abundance ratio and the \citet{sheffer2008ultraviolet} relation, we derive the OH abundance using CH as a surrogate for \molh to be $X_{\text{OH}} = N$(OH)/$N$(\molh) = $(1.09 \pm 0.27)\times10^{-7}$. This result is consistent with the column density ratios presented by \citet{liszt2002comparative}, and is in the range of theoretical predictions presented by \citet{2014ApJ...787...44A}. When using a $N$(CH)/$N$(H$_{2}$) ratio of $\sim 4.97\times10^{-8}$ as derived by \citet{weselak2019relation}, we find the corresponding OH abundance to be $X$(OH) = $1.56\times10^{-7}$.

\subsection{Radial distribution of CH}\label{subsec:radial_dist}
\begin{figure}
    \centering
    \includegraphics[width=0.495\textwidth]{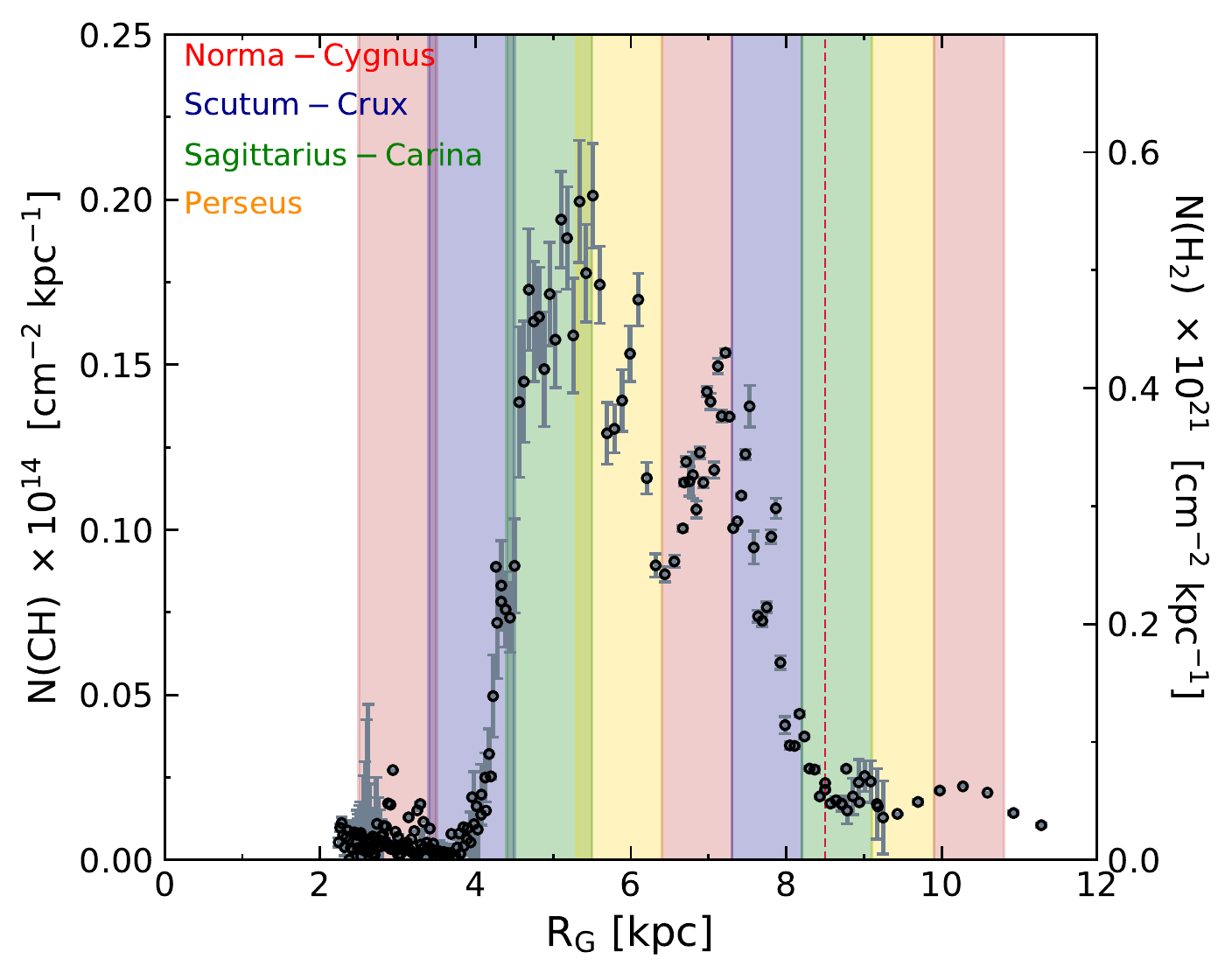}
    \caption{Radial distribution of CH column densities toward lines of sight probed in this study except SgrB2(M). The column density values for each distance interval are averages over the underlying distribution. The secondary y-axis display $N(\text{H}_{2})$ values computed using the \citet{sheffer2008ultraviolet} relation. The dashed red vertical line indicates the galactocentric distance to the Sun. The spiral arm locations typical for a fourth quadrant source, are plotted with spiral arm widths of 0.85~kpc.}
    \label{fig:galactic_distribution}
\end{figure}

 The azimuthally-averaged CH column densities obtained toward all the sightlines in this study, except SgrB2(M), are plotted as a function of galactocentric distances, $R_\text{G}$, in Fig.~\ref{fig:galactic_distribution}. The galactocentric distances to the different spiral arm features were computed assuming a flat rotation curve, with the distance of the Sun from the Galactic centre given by R$_{0}$ = 8.3~kpc, and an orbital velocity of $\Theta_{0}$ = 240~km~s$^{-1}$ of the Sun, with respect to the Galactic centre \citep{reid2014trigonometric}. The uncertainties in the CH column densities are typically of the order of a few $10^{12}\,$cm$^{-2}$(km~s$^{-1}$)$^{-1}$, and are too small to be visible in some parts of the plot. This analysis is limited to a small number of strong background continuum sources, distributed at low Galactic latitudes, which lie close to the Galactic plane.

We find that the CH column densities peak at galactocentric distances between $4$ and $8\,$kpc. LOSs within the $l$ = 0-180$^{\circ}$ range dominate contributions toward the first column density peak observed around 5$\,$kpc, while the second peak near $7\,$kpc originates from LOSs in the $l$ = 180-360$^{\circ}$ range. The column density peak near $5\,$kpc is associated with absorption arising from a combination of the Perseus and Norma spiral-arm crossings, and the $7\,$kpc peak may be attributed to the intersection of the Scutum-Crux spiral-arm crossing along the sightlines.
The higher observed column densities peaking near these spiral arms can be attributed to the larger (mass) molecular content that is present at these galactocentric radii. Owing to the larger error bars, the significance of a third peak near 9$\,$kpc is unclear. However, it is in line with contributions that may arise from the Scutum- and Sagittarius-arm far side crossings. Due to a lack of data points at galactocentric distances smaller than 2$\,$kpc, the nature of the CH molecular gas distribution there remains unknown. We do not include contributions from the sightline toward SgrB2(M) in this analysis, because in the Galactic centre direction, it is difficult to assign the CH absorption to different components based on their LSR velocities. 

A similar dual-peaked radial distribution profile has been reported by both \citet{pineda2013herschel}, and \citet{velusamy2014origin} for the cold neutral medium (CNM) component of \cii  emission. \cii  is a proficient tracer that is sensitive to various evolutionary stages of the ISM, but by taking advantage of differing densities, \citet{pineda2013herschel} were able to decompose contributions of the \cii emissivity from the different phases of the ISM, excluding the warm ionised medium (WIM) for sources in the Galactic plane, $b$ = 0$^{\circ}$. Despite the smaller sample size, our results broadly resemble the shape of the \cii integrated intensity distribution profile.

The framework of carbon chemistry in the diffuse regions, which is initiated by the radiative association of the ground vibrational state of \molh with ionised carbon, explains the observed correspondence between CH and \cii radial profiles \citep{Gerin2016}.
Peaking at the same $R_\text{G}$ values as those of H{\small I} (see Fig.~12 in \citet{pineda2013herschel}), suggests that CH likely traces the more diffuse regions of the CNM between \cii and the extended H{\small I} emission. Upon comparing the distribution of \molh column densities derived from CH (using the \citet{sheffer2008ultraviolet} relation) with those derived from both CO (CO-traced \molh gas) and \cii (CO-dark \molh gas) by \citet{pineda2013herschel} (their Fig.~18) we find that: (1) at R$_\text{G}$ $<$ 2~kpc, we cannot comment on the behaviour of the distribution due to the lack of data points, (2) 2~kpc $<$ R$_\text{G}$ $<$ 8~kpc, the CH-traced \molh column density is comparable to the [C{\small II}]-traced \molh column density, while being smaller than the CO-traced \molh column density and, (3) R$_\text{G}$ $>$ 8~kpc, despite attaining a constant level at these distances, similar to the CO-dark \molh gas traced by {[C\small II]}, the column density distribution of \molh obtained using CH is lower than that traced by {[C\small II]}. 

The general agreement between the column density of the CO-dark gas traced by [C{\small II}] and the \molh column densities traced by CH at galactocentric distances $<8\,$kpc, is a good indicator of the usefulness of CH as a tracer for the CO-dark component of molecular gas.

\section{Conclusions}\label{sec:summary}
In this paper, we presented the analysis of SOFIA/GREAT observations of the $N, J = 2, 3/2 \rightarrow 1, 1/2$ transitions of CH detected in absorption toward strong, FIR-bright continuum sources by using the Wiener filter algorithm. Although it has previously not been used as a conventional tool for spectral line analysis, we find the WF deconvolution algorithm to be a self-consistent concept that sufficiently alleviates biases manifested in spectra with close hyperfine spacing, provided that the transitions are optically thin. The absorption features along each LOS are broadly classified into contributions arising from different spiral arm and inter-arm regions based on the spiral arm model of the Milky Way presented by \citet{reid2014trigonometric}. Using the WF as the basis for our spectral analysis, we derived a linear relationship between CH and OH, $N$(OH)/$N$(CH) = $3.85 \pm 0.15$ that is consistent with previous work. Upon increasing the number of sightlines in this analysis by including those studied by \citet{weselak2010relation}, we derived a revised relationship with a $N$(OH)/$N$(CH) = 3.14 $\pm$ 0.49. By subsequently using CH as a surrogate for tracing H$_{2}$, we derived an OH abundance of $X$(OH) = ($1.09 \pm 0.27$)$\times 10^{-7}$. It is important to note that this result pivots on the validity of the $N$(CH)-$N$(\molh) relation used. We find the radial distribution of the CH abundances of our sample to peak between $4$ and $8\,$kpc. In addition to the tight correlation that exists between the CH and \molh established by \citet{sheffer2008ultraviolet}, the likeness of the CH abundance distribution to that of \cii, a well known tracer, of the CO-dark \molh gas, lends credence to its use as tracer for \molh in these regions. 
Furthermore, this work may initiate follow-up observational studies, investigating a larger sample of sources covering a wide range of physical conditions at varying galactocentric distances to study the Galactic distribution of CH. This will reinforce its use as a tracer for \molh in regions where hydrogen is molecular, and carbon may be both neutral or ionised, but not traced by CO, over Galactic scales.

\begin{acknowledgements}
SOFIA Science Mission Operations is jointly operated by the Universities Space Research Association, Inc., under NASA contract NAS2-97001, and the Deutsches SOFIA Institut under DLR contract 50 OK 0901 and 50 OK 1301 to the University of Stuttgart. We are GREATful to the SOFIA operations team for their help and support through out the course of the observations and after. A. Jacob is a fellow of the  International Max-Planck-Research School (IMPRS) for Astronomy and Astrophysics at the Universities  of Bonn and Cologne. We also thank Carsten K\"{o}nig and Dario Colombo for their helpful discussions on identifying the different spiral-arm and inter-arm LOS crossings. M.-Y.L. was partially funded through the sub-project A6 of the Collaborative Research Council 956, funded by the Deutsche Forschungsgemeinschaft (DFG).
\end{acknowledgements}

\bibliographystyle{aa.bst}         
\bibliography{ref.bib}

\begin{appendix}

\section{Convolution model}\label{appendix:convolution_model}
\begin{figure*}
\includegraphics[width =1.01\textwidth]{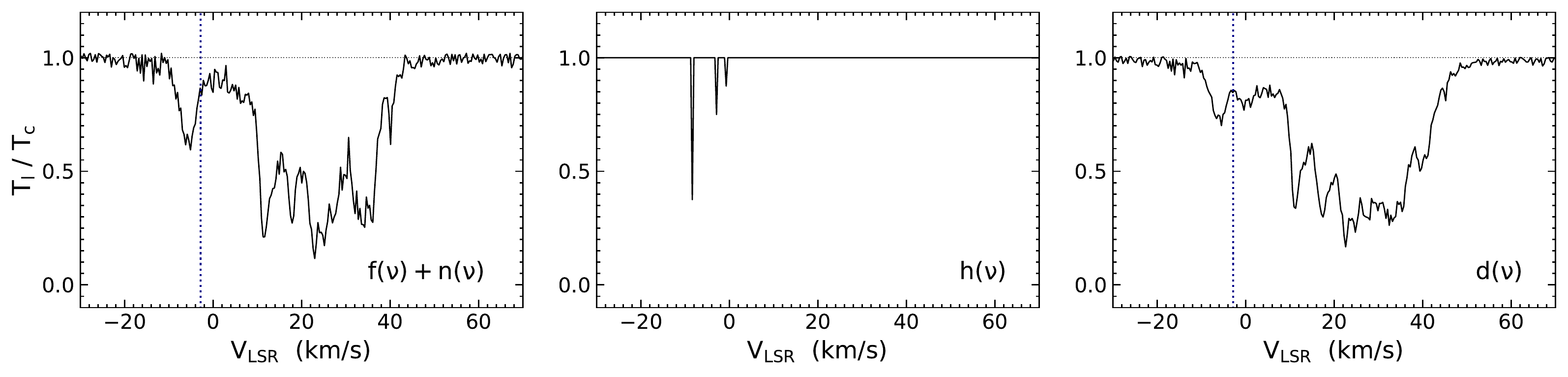}
\caption{Convolution model illustrated for LOS toward AGAL010.624$-$00.384. The input signal with additive noise or the original (deconvolved) spectrum (left) is convolved with the normalized HFS weights (center) to produce the resulting convolved spectrum or the observed noise-degraded spectrum (right).}
\label{fig:conv_mdl}
\end{figure*}

Figure~\ref{fig:conv_mdl} presents an example that showcases the effects of HFS splitting on astrophysical spectra. The deviations between the observed spectrum and noise-degraded spectrum (resulting from the convolution model) computed using residuals, vary on average between $-$0.03 and +0.02.

\section{Assessment of the WF fit}\label{appendix:residuals}

\begin{figure}
%\sidecaption
\includegraphics[width=0.5\textwidth]{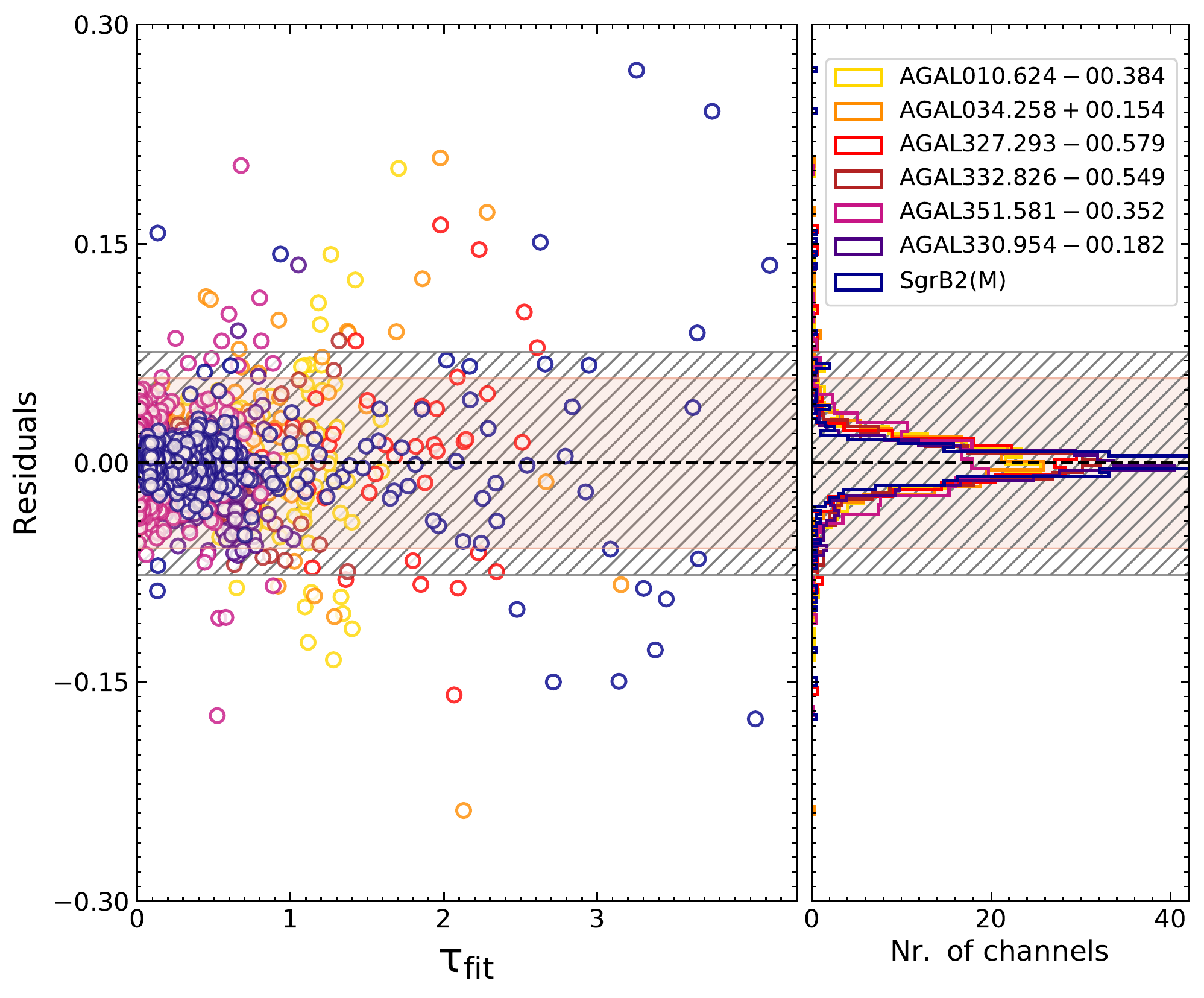}
\caption{Left panel: optical depth residuals vs. WF fit. The different coloured markers correspond to residuals computed from the WF fits to the different sources. Right panel: distribution of residuals corresponding to different sources. They follow approximate normal distributions centred slightly off-zero around +0.001 for all seven sources. The grey (hatched) and pink regions bound the $99\, \%$ and $95\, \%$ confidence intervals, respectively.  }
\label{fig:residuals}
\end{figure}

In this appendix, we discuss the quality of the WF fit prior to deconvolution. This analysis aims to curb the propagation of biases introduced from an inadequate fit into any derived quantities. For this reason, we carry out a residual analysis to test the validity of the WF fitting model and its underlying assumptions. The ``residuals'' of a fit may be viewed as deviations in the values predicted by the model from what is observed. Given that the Wiener filter itself is derived by minimising the mean square error between the data and the model (Sect.~\ref{subsec:WFD}), one would not expect large discrepancies. Figure~\ref{fig:residuals} displays the combined results of the residuals obtained for each source. 

  The left panel shows no clear pattern or structure, with the residuals randomly scattered around zero. The lack of any uniform structure reveals that our assumptions of a linear system with an independent noise signal is maintained. 
  From the right panel, we see that the residuals are normally distributed for each source, with a mean of $0.001$, and standard deviation of 0.024. Only 2.4 and 3.7$\, \%$ of all the residual points lie outside the 99 and 95$\, \%$ confidence intervals. These outliers can be attributed to small variations in the power spectrum of the additive noise that lead to an over-(positive-valued residual) or under-(negative-valued residual) estimation of the true noise level. Such inaccuracies in the estimated noise are more pronounced at the spectral peaks, and correspond to the outliers seen in Fig.~\ref{fig:residuals}.
  Overall, the residuals validate our fit and confirm our assumptions by reproducing independent errors that are not serially-correlated with a near constant variance and normal distribution.

\section{Computational efficiency of the WF algorithm}\label{appendix:forward}

From the mathematical formulation presented in Sect.~\ref{subsec:WFD}, the WF can be interpreted as a kernel acting on the inverse filter. If the signal happens to be much stronger than the noise, then (S/N)$^{-1}\approx~0$, in which case the WF kernel reduces to $H^{-1}(\tilde{\nu})$ or the inverse filter. Alternatively, if the signal is weak in comparison to the noise, then (S/N)$^{-1} \rightarrow \infty$ and $W(\tilde{\nu}) \rightarrow 0$ and the WF attenuates contributions from higher noise levels which characterises it as a bandpass filter. 

WF deconvolution, while efficient in its performance, is still an approximate method, in the sense that its main drawbacks arise from the MSE constraint, which is also, ironically, what sets it apart form other filters. The process of aptly minimising the overall MSE carries out inverse filtering by accounting for additive noise. However, since the human eye is willing to accept more noise than allowed for by the filter (provided that it is spatially associated with the spectral profile), the restored spectra might seem unpleasantly smooth. This is because the trade-off between filtering and noise smoothing holds true only up to a second order in precision, meaning the required levels of degradation are not always achieved when characterising the signal. This leads to deviations between the obtained results and the true optimum. Hence the main performance limitation arises form the finite variance in the noise at any given channel and its treatment. These effects are seen in the form of small fluctuations in the WF fit (as discussed in Appendix~\ref{appendix:residuals}), particularly in regions of the spectra where the S/N is very large. One could then turn toward using this response as a respectable initial guess, and repeating the same filtering process over multiple iterations in order to improve the restored spectrum. This iterative line of thought is futile, because the scheme converges as H($\tilde{\nu}$) $\rightarrow$ 0. The non-iterative nature of the WF is in fact one of its biggest advantages, as it is less time-consuming, and, additionally, aids in any subsequent Monte Carlo-based error estimates.

\begin{figure}
\includegraphics[width=0.5\textwidth]{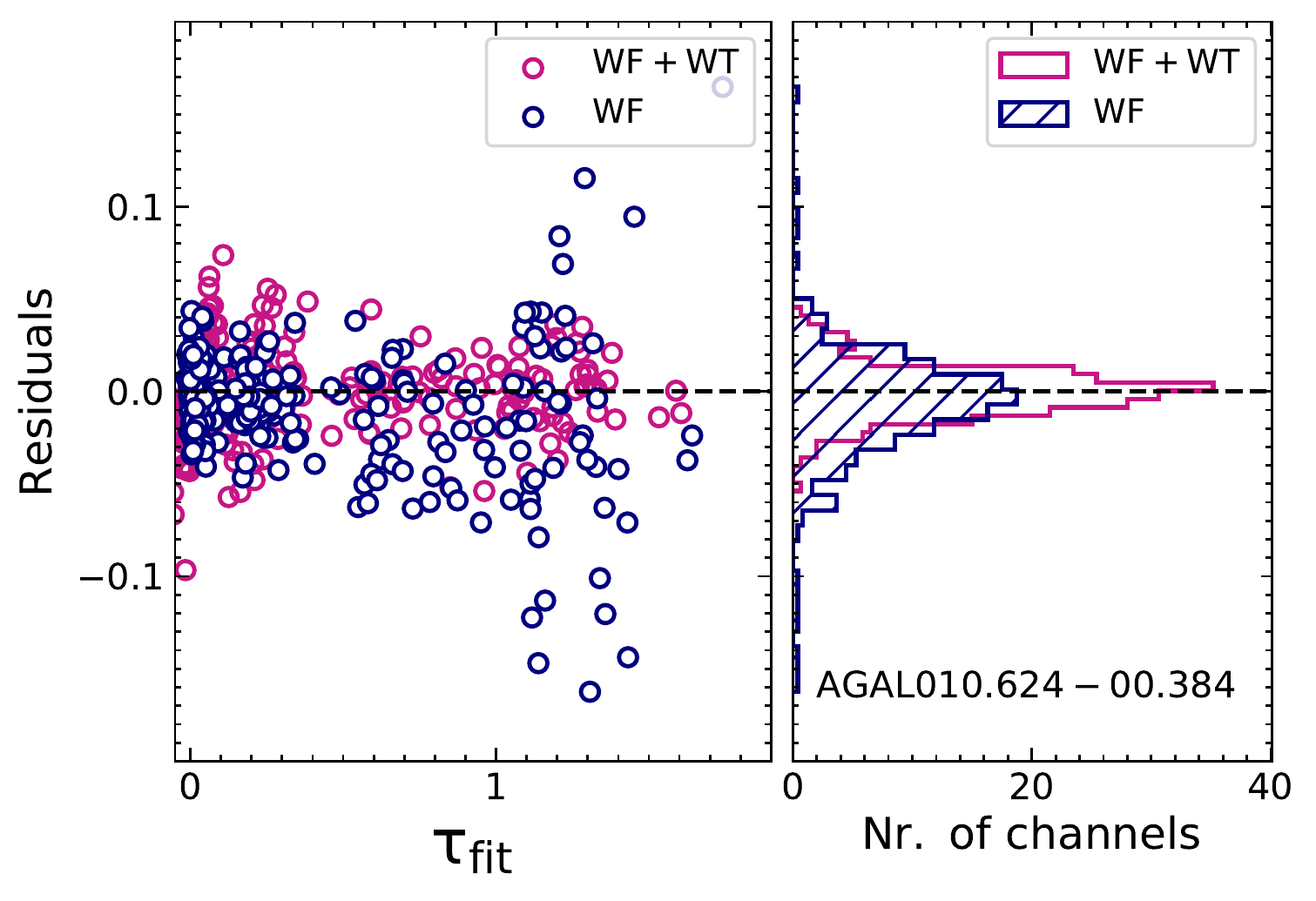}
\caption{Left: scatter plots of residuals obtained from the WF fit and WF + WT (ForWARD) algorithm against their fits in dark blue and magenta, respectively. Right: weighted distributions of residuals from the WF fit alone (blue hatched region) and the WF + WT (magenta).}
\label{fig:wavelet}
\end{figure}

The wavelet transform (WT) is another method used to quantify the degree of noise attenuation. Based on the quantization of a signal in the form of wavelets, the WT has the advantage that it characterises the local behaviour of a signal while maintaining all temporal information, unlike the FT. The Fourier-wavelet regularisation deconvolution (ForWARD) algorithm developed by \citet{neelamani2004forward} is one such technique that combines deconvolution in the Fourier domain with noise suppression in the wavelet domain. In this approach, the WF deconvolution product is first decomposed in the wavelet domain into a scaling function, which acts as a low-pass filter and a wavelet function or a band-pass filter, which in turn constitutes each wavelet level. A second step of wavelet de-noising was added to our existing WF algorithm, in order to improve the robustness and efficiency of this method. Once again, we visualise the fit through its residuals as shown in Fig.~\ref{fig:wavelet} toward a single source AGAL010.624$-00$.384. 

In general, both sets of residuals are scattered randomly and centred around zero. The outliers in the WF fit are curbed when it is followed by the use of the WT, giving rise to a less variant or narrower distribution. However, the wavelet domain analysis is more suited to transient or time-varying signals, whereas our spectra are time independent. A subtle problem that this creates in a time-invariant system is that the wavelet domain is sparse (reduced number of data points) because of the wavelet decomposition of the spectral signal, and remains sparse over consecutive iterations. This makes it difficult to conclude whether the WT actually takes into account the outliers, or whether their effects are averaged out in the process of forming wavelets. While the addition of the WT to the WF deconvolution makes it more robust and rigorous, it has a longer computational time, and application to time-invariant systems makes it non-ideal.

\section{Impact of using a single excitation temperature on derived column densities} \label{appendix:impact_excitationtemp}
\begin{figure}
\vspace{-0.05cm}
    \centering
    \includegraphics[width=0.443\textwidth]{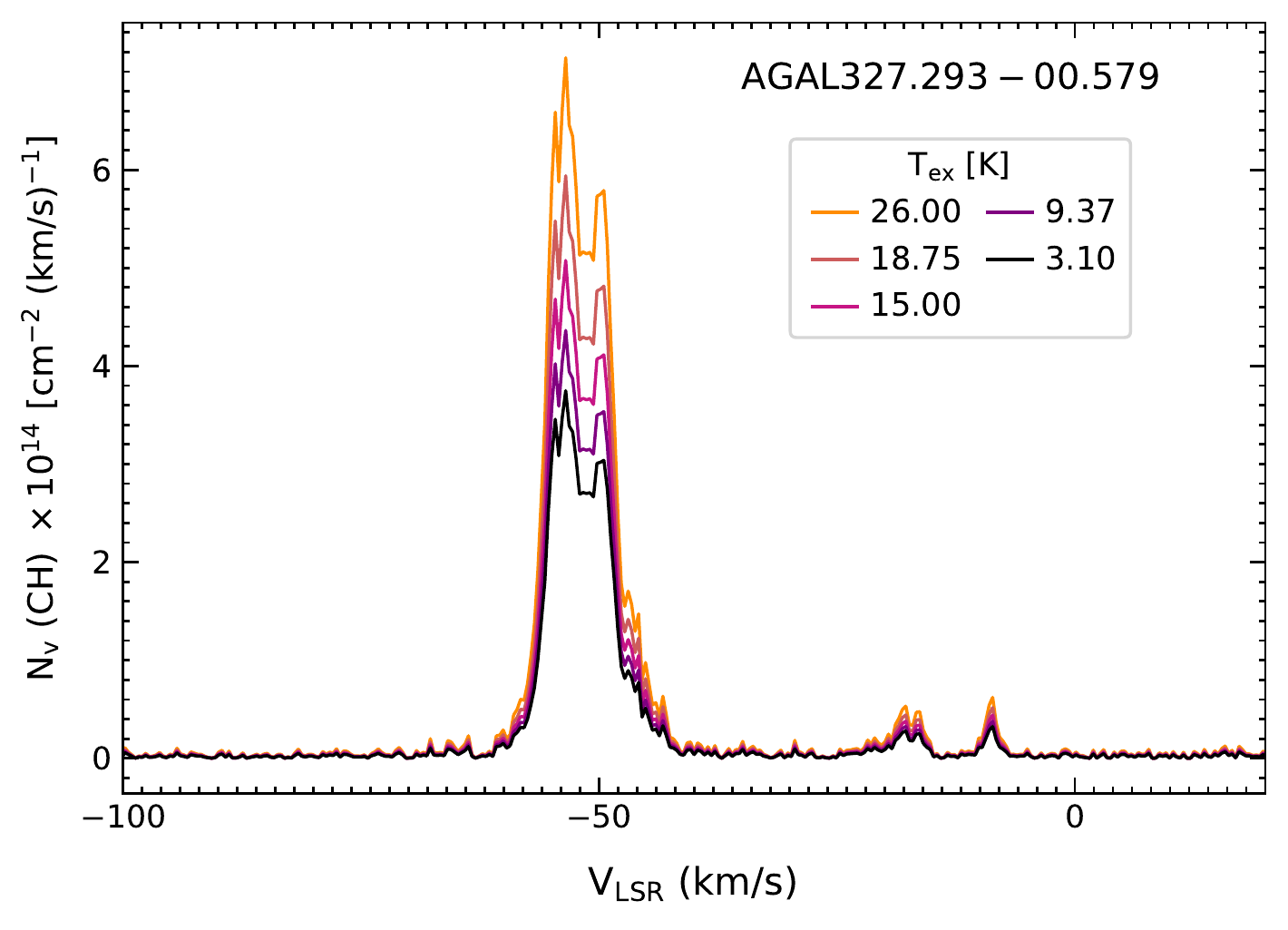}
    \vspace{-0.14cm}
    \caption{Variation in WF deconvolved column density profile toward AGAL327.293$-00$.579 at varying values of excitation temperature between 3.1 and 26$\,$K.}
    \label{fig:tex_impact}
\end{figure}

The column densities presented in Table.~\ref{tab:coldens} are derived using the WF deconvolution by assuming a single excitation temperature of $3.1\,$K for all components. Since the critical densities, $n_\text{crit}$\footnote{The critical densities for CH are computed using rate coefficients calculated by \citet{dagdigian2018hyperfine} and Einstein A coefficients as given in Table.~\ref{tab:freq}.} of the CH transitions presented in this work ($\sim 2.4\times10^{9}\, \text{cm}^{-3}$, assuming a gas temperature value of $50\,$K) are several orders of magnitude higher than the typical conditions that prevail over the foreground LOS clouds, almost all the CH molecules are expected to be in the ground state. Furthermore, the large values of $n_\text{crit}$ imply that population of any rotational level above the ground state of CH must arise from radiative excitation processes rather than collisional excitation, as shown by \citet{black1994energy}. As mentioned in Sect.~\ref{subsec:WFD_lineprofile}, radiative excitation is unlikely for CH molecules residing in locations that correspond to velocity intervals associated with the diffuse and the translucent regions. Using higher values for the excitation temperature, we model the absorption features associated with the molecular environment of the bright FIR continuum sources (Fig.~\ref{fig:tex_impact}). Changes in the excitation temperature between 3.1 and 10$\,$K result in uncertainties in the calculated column densities that are less than 15$\%$, while this margin increases to nearly 50$\%$ as $T_\text{ex} \rightarrow 30\,$K. While a proper estimation of the excitation temperature requires a thorough radiative transfer description across the envelope of the continuum source, here, we merely illustrate how the calculated column densities are affected by  the assumption of a single excitation temperature.  

\section{Additional tables}\label{appendix:additional}
\begin{table}[H]
\caption{Spectroscopic parameters of other studied transitions.}
\begin{center}

\begin{tabular}{cccccc}
\hline \hline
Species & \multicolumn{2}{c}{Transition} &  Frequency & $A_\text{E}$  &  $E_{u}$ \\
& $J$ & $Parity,  F$&  [GHz]   & [s$^{-1}$]&     [K]      \\
\toprule
CH &  3/2 $\rightarrow$ 1/2  & $^{-}1 \rightarrow 1^{+}$ & $\,\,$532.7216 & $\,\,$0.020 & $\,\,$25.76 \\
     &               & $^{-}1 \rightarrow 0^{+}$ & $\,\,$532.7239 & $\,\,$0.062   \\
	&			& $^{-}2 \rightarrow 1^{+}$ & $\,\,$532.7933 & $\,\,$0.041  \\\midrule  
	
OH & 5/2 $\rightarrow$ 3/2& $^{+}2 \rightarrow 2^{-}$ & 2514.2987 & $\,\,\,$1.367
 & 120.75 \\
      &               & $^{+}2 \rightarrow 3^{-}$ & 2514.3167 & 13.678   \\
	& &			$^{+}1 \rightarrow 2^{-}$ & 2514.3532 & 12.310 \\
                    \bottomrule

\label{tab:freqothers}
\end{tabular}
\end{center}
\tablefoot{The spectroscopic data for other studied CH and OH transitions - quantum numbers, frequencies, Einstein-A coefficients and upper level energies were taken from the Cologne Database for Molecular Spectroscopy ~\citep{muller2001cologne}.}
\end{table}

\section{Correlation plots for individual sources }\label{appendix:correlations}
This Appendix showcases the $N$(OH)-$N$(CH) correlations toward the individual sources. 
\begin{figure*}
    \centering
    \includegraphics[width=1\textwidth]{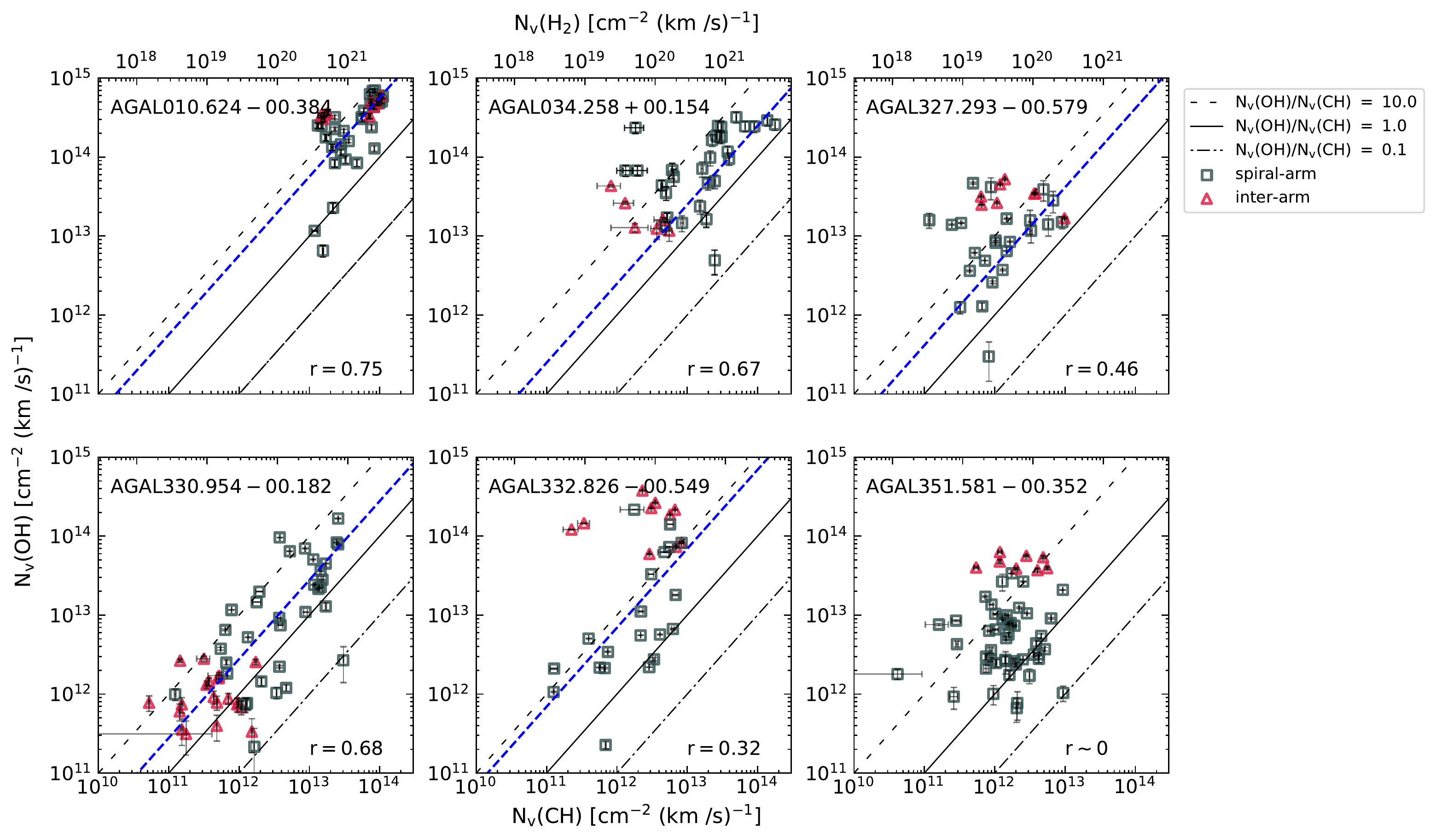}
    \caption{Correlation between CH and OH column densities derived using the WF deconvolution algorithm toward (clockwise from top left) AGAL010.624$-00$.384, AGAL034.258+00.154, AGAL327.293$-00$.579, AGAL330.954$-00$.182, AGAL332.826$-00$.549 and AGAL351.58$-00$.352. The different markers indicate contributions from spiral-arm (in blue), and inter-arm gas (in red). The linear regression is shown by the blue dashed line along with $N$(OH)/$N$(CH) ratios of 0.1, one and 10. Secondary x-axis represents \molh column densities obtained using the CH/\molh abundance ratio given by \citet{sheffer2008ultraviolet}. The corresponding Pearson's correlation coefficients are given in the lower-right corners.}   \label{fig:individual_correlation}
\end{figure*}

\begin{table}[H]
    \centering
    \caption{Synopsis of $N$(OH)/$N$(CH) regression toward individual sources, except G351.581$-00$.352.}
    \begin{tabular}{cccc}
    \hline \hline
         Source & $N$(OH)/$N$(CH) & Pearson's   &$N$(OH)/$N$(H$_{2}$) \\
                &              &   r-value& $(\times 10^{-7})$ \\ \hline
         AGAL010.624$-00$.384  &  $5.66\, \pm0.06$ & 0.75&  $1.98 $\\
         AGAL034.258+00.154  &  $2.48\, \pm0.17$ & 0.67 & $0.87 $ \\
         AGAL327.293$-00$.579 &  $4.13\, \pm0.11$ &  0.46 & $1.44 $ \\
         AGAL330.954$-00$.182 &  $2.78\, \pm0.07$ &  0.68 & $0.97 $ \\
         AGAL332.826$-00$.549 &  $9.70\, \pm0.48$ &  0.32 & $3.40 $ \\
          \hline
    \end{tabular}
  
    \label{tab:individual_correlations}
\end{table}
\end{appendix}

\end{document}